\documentclass[twocolumn,showpacs,preprintnumbers,amsmath,amssymb]{revtex4-1}

\usepackage{graphicx}
\usepackage{dcolumn}
\usepackage{bm}
\usepackage{amsmath}
\usepackage{amssymb}
\usepackage{multirow}

\DeclareMathOperator{\sgn}{sgn}

\begin{document}

\preprint{}

\title{Lattice equations and dielectric function for optical lattice vibrations in monolayer transition metal dichalcogenides}

\author{J.-Z. Zhang}
 \email{phyjzzhang@jlu.edu.cn}
 \affiliation{School of Physics, Jilin University, Changchun, 130012, China.
}

\date{\today}
\begin{abstract}
Using a microscopic dipole lattice model including electronic polarization (EP) of ions and local field effects (LFEs) self-consistently, two sets of three equations are deduced for long wavelength in-plane and out-of-plane optical lattice vibrations in monolayer (ML) transition metal dichalcogenides (TMDs). Expressions for the lattice vibrational energy density are obtained for the two-dimensional (2D) crystals. The linear coefficients of the lattice equations can be determined by first-principles calculation, making these equations very useful for studying the lattice dynamical properties. 
The two pairs of equations describing the polar vibrations have the same forms as those for ML  hexagonal BN and also resemble Huang's equations for bulk [Proc. Roy. Soc. A {\bf 208}, 352 (1951)]. Each pair of the equations are solved simultaneously with the equation of electrostatics, and 
explicit expressions are obtained for the in-plane longitudinal and transverse optical (LO and TO) modes and out-of-plane optical (ZO) modes. 
The LO phonon dispersion relation is identical to the analytical expression of Sohier {\it et al}. [Nano Lett. {\bf 17}, 3758 (2017)], and it expresses the degeneracy of the LO and TO modes at $\Gamma$ and their splitting at finite wavevectors due to the long-range macroscopic field. All the optical phonon branches except for LO are nondispersive at the long wavelengths. 
A 2D longitudinal lattice dielectric function $\epsilon(k,\omega)$ is deduced, allowing one to rederive the LO phonon dispersion simply from $\epsilon(k,\omega)=0$. 
A 2D Lyddane--Sachs--Teller relation  and a frequency--susceptibility relation are obtained for in-plane and out-of-plane vibrations, respectively, through which the phonon frequencies are related to the 2D dielectric functions or susceptibilities.     
The explicit expressions are applied to calculate various dynamical properties when knowing three first-principles calculated parameters,  and   
the ionic EP and LFEs are studied thoroughly. 
 To evaluate the LFEs the unit-cell atomic polarizability is introduced, which is found to be in an interval using the 2D Clausius-Mossotti relation.  
The EP and LFEs should be included simultaneously; otherwise, neglecting either or both causes substantial underestimates to key dynamical quantities, such as the Born charge, the static and high-frequency dielectric susceptibilities and the LO phonon dispersion.

\end{abstract}

\maketitle

\section{Introduction}
Interest in two-dimensional (2D) materials such as monolayer (ML) transition metal dichalcogenides (TMDs) has surged in recent years due to their potential for novel electronic and optoelectronic devices  \cite{Geim:2013,Wangqh:2012,Makf1:2012,Fiori:2014}. 
 The 
ML TMDs such as MoS$_2$ are 2D polar crystals with a honeycomb lattice structure, in which the polar lattice vibrations occur together with the nonpolar vibrations.  Particularly interesting are the long wavelength polar optical vibrations; for instance, the longitudinal optical (LO) vibrations can generate a macroscopic electric field that strongly couples to electrons, and this makes the polar electron-phonon interaction dominate the electron transport processes in 2D TMDs. 
The 2D optical phonons have attracted great interest as they play a key role in carrier transport \cite{Kaasbjerg:2012,Lixd:2013,Danovich:2017},  piezoelectricity \cite{Wuw:2014,Zhuh:2014,Michel:2017}, valley depolarization \cite{Miller:2019} and homogeneous broadening of excitonic luminescence \cite{Moody:2015,Jakubczyk:2016} in ML TMDs. Hence, an in-depth knowledge of the 2D lattice vibrations is very useful for the study of the electronic and optical properties of ML TMDs. 


A generally used method for phonon calculations is the {\it numerical} solution of the dynamical matrix \cite{Born:1954}, and with it the phonon spectra in the entire Brillouin zone can be obtained. For isotropic bulk crystals the macroscopic theory developed by Huang allows for an {\it analytical} description of the long optical lattice vibrations by a pair of equations, $\ddot{\mathbf{w}}=b_{11}\mathbf{w}+b_{12}\mathbf{E}$,  $\mathbf{P}=b_{21}\mathbf{w}+b_{22}\mathbf{E}$ ($b_{12}=b_{21}$) \cite{Huang:1951,Huang:1951a,Born:1954}, where $\mathbf{w}$ is used for describing the optical displacement of the lattice, and $\mathbf{E}$ and $\mathbf{P}$ are the electric field and dielectric polarization as in Maxwell's equations, both being {\it macroscopic} quantities, averaged over a lattice {\it cell} \cite{Born:1954,Kittel:2004}. Compared to the Hooke's law formula, the right-hand side of the equation of motion has an extra term due to the electric force, while in the polarization equation there is a lattice displacement contribution as well as the usual field-induced polarization. Microscopically, the field acting on any particular ion is a {\it local} field (Lorentz field), i.e., the total field of all the ions except for the ion itself, but it is eliminated by means of the {\it Lorentz relation} \cite{Born:1954,Kittel:2004}, which is essential in Huang's deduction. The electronic polarization (EP) of ions is also accounted for in Huang's model. Using Huang's equations not only the long wavelength optical modes but also the lattice dielectric function is obtained, and further the well-known Lyddane--Sachs--Teller (LST) relation \cite{Lyddane:1941} is rederived \cite{Born:1954}. 

Like in bulk polar crystals, 
the long wavelength lattice vibrations in a 2D polar crystal are closely connected with the macroscopic and local fields in the crystal. In an isotropic bulk crystal no macroscopic field occurs with the transverse optical (TO) vibrations but the LO vibrations generate a finite long-range macroscopic field, leading to a higher LO phonon frequency and LO-TO {\it splitting}  \cite{Lyddane:1941}. 
In a 2D crystal, the lattice may polarize according to the in-plane or out-of-plane lattice motion, giving rise to the macroscopic fields that differ for the in-plane and out-of-plane polarization \cite{Zhang:2019b}. While there is no macroscopic field accompanying the TO vibrations, like in the bulk, the macroscopic field of the LO vibrations differs substantially from that in the bulk, as it vanishes in the long wavelength limit,  leading to the {\it degeneracy} of the LO and TO modes at the $\Gamma$ point \cite{Sanchez:2002}. Further, the macroscopic field due to the out-of-plane vibrations becomes extremely large close to the atomic layers as the lattice polarization is of only atomic scale in the out-of-plane direction $z$ \cite{Zhang:2019b}.  Similarly, the local field in the 2D crystal at an ion site strongly depends on the lattice motion, making the local fields differ for the in-plane and out-of-plane lattice vibrations \cite{Mikhailov:2013}. In fact these are very strong local fields \cite{Mikhailov:2013,Zhang:2019b}, which is different from the case of the bulk where the local fields on the ions are quite weak and largely cancelled \cite{Born:1954}.  

As the wavevector is parallel to the layer planes, the lattice vibrations of 2D crystals comprise the in-plane and out-of-plane vibrational modes. In ML hBN there are three branches of optical modes, all of which are polar modes. 
With one positive ion and two same negative ions per unit cell, a ML TMD has three branches of nonpolar optical modes corresponding to the relative motion the two negative ions, as well as three branches of polar optical modes like those of ML hBN, which are connected with the contrary motion of the positive and negative ions.      
Specifically these optical branches are composed of the in-plane nonpolar LO and TO modes (labeled LO$_1$, TO$_1$), the in-plane polar LO and TO modes (labeled LO$_2$, TO$_2$), the out-of-plane nonpolar modes, ZO$_1$, and the out-of-plane polar modes ZO$_2$. At the $\Gamma$ point of the Brillouin zone the LO$_1$ and TO$_1$ are degenerate and denoted as $E''$ by symmetry consideration, the doubly degenerate LO$_2$ and TO$_2$ modes are denoted by  
$E'$, and the out-of-plane ZO$_1$ and ZO$_2$ modes are denoted as $A_1$ and $A_2''$, respectively \cite{Molinasan:2011}. 
$A_2''$ is an infrared active mode, while $A_1$ and $E''$ are Raman-active modes, and the $E'$ modes are both infrared and Raman active.    
The  $A_1$ and $E'$ modes have been measured by Raman spectroscopy for a number of ML TMDs \cite{Leec:2010,Tongay:2012,Lanzillo:2013,Rice:2013,Saito:2016,Lux:2016}. 

Phonon dispersion relations of ML TMDs have been studied based on first-principles calculations \cite{Molinasan:2011,Ataca:2011,Sohier:2017}. 
Phonon modes calculated from first principles have been used to study mechanical stress-strain relations \cite{Lits:2012}, electron transport properties \cite{Kaasbjerg:2012,Lixd:2013}, thermal conductivity \cite{Cai:2014} and thermal expansion \cite{Cai:2014,Huanglf:2014,Cakir:2014} and piezoelectricity \cite{Michel:2017} in ML TMDs. 
These calculated phonon spectra show several common features on the {\it optical}  branches: 
(i) apart from the degeneracy of the in-plane nonpolar modes LO$_1$ and TO$_1$, the 
 polar modes LO$_2$ and TO$_2$ are also degenerate at the $\Gamma$ point; 
(ii) the LO$_2$ and TO$_2$ modes split at finite wavevectors, and the slow increase of LO phonon frequency leads to a small overbending (i.e., the maximum frequency occurs at a wavevector away from $\Gamma$ \cite{Sohier:2017}) in the LO phonon dispersion compared to ML hBN \cite{Topsakal:2009,Sahin:2009,Sohier:2017};  
(iii) the ZO$_1$ nonpolar branch displays a nondispersive behaviour in the Brillouin zone. 
While most of the theoretical studies are based on first-principles calculations, there is a lack of analytical approaches and there are no equations of motion to describe the long optical lattice vibrations in ML TMDs. 
Using a simple model, Sohier {\it et al}. have recently derived an LO phonon dispersion relation \cite{Sohier:2017} for the 2D polar crystals, including hBN and TMDs, and the parameters in the dispersion relation can be calculated from first-principles to  include ionic EP and local field effects (LFEs). The dispersion relation can well describe the dispersion of the numerically calculated polar optical modes \cite{Sohier:2017}.

Recently we have deduced lattice equations for the long wavelength optical vibrations in ML hBN using a microscopic model, and then derived optical phonon dispersion relations in which the LO phonon dispersion relation is identical to the expression of Sohier {\it et al}. \cite{Sohier:2017}. Here one purpose of this study is to extend our approaches to analytically tackle the lattice vibrations of ML TMDs. To formulate a macroscopic theory of the optical vibrations, both {\it macroscopic quantities}, i.e., the macroscopic field to drive the lattice motion and the 2D dielectric polarization are needed \cite{Zhang:2019b}. They both need to be defined at first, and a digestion of the results on ML hBN \cite{Zhang:2019b} will give us some guidance.  
 The macroscopic field act on all the ions of a lattice cell so in principle it should have the {\it same} value for all these ions \cite{Born:1954,Kittel:2004}. 
As ML hBN has only a single atomic layer, one of course uses the macroscopic field in the ML plane, and also defines the 2D lattice polarization straightforward  \cite{Zhang:2019b}. 
Unlike ML hBN, however 
a ML TMD consists of three atomic layers at {\it different} $z_i$ ($i=1, 2, \bar{2}$) positions and has a finite thickness. As 
each of the three layers resembles a hBN monolayer, using the macroscopic field result of Ref.\cite{Zhang:2019b}  each in-plane polarized layer produces a macroscopic field that varies with $z$ according to $e^{-k\lvert z-z_i\rvert}$ ($k$ is a small wavevector), which varies very slowly inside the ML because for any $z$ in the ML $k\lvert z-z_i\rvert$ is a small quantity, $k\lvert z-z_i\rvert\ll 1$. Therefore we use the macroscopic field at the {\it central} layer $z_1$  (positive ion site) to represent the macroscopic field acting on all the three layers (all the three ions of a cell).  
Further, this shows that the different $z_i$ positions of the layers has become unimportant and the 2D TMD behaves like a single layer crystal to the long lattice waves. Therefore we can define a 2D macroscopic polarization from the microscopic electric dipole density, by averaging its 
 microscopic $\delta(z-z_i)$ distributions at the three layers (Sec. II below). For the out-of-plane polarization, 
the electric fields of the three polarized atomic layers are equally divergent according to $\delta(z-z_i)$ \cite{Zhang:2019b} so the averaging treatment can be also used to handle the out-of-plane lattice motion.

Having clarified these points regarding the macroscopic field and polarization, in this paper, therefore we can extend the approaches of the previous work \cite{Zhang:2019b} to study systematically the long wavelength optical vibrations in ML TMDs. 
We deduce lattice equations involving both macroscopic quantities using a microscopic model that accounts for ionic EP and LFEs self-consistently. Then we solve the  lattice equations simultaneously with the equation of electrostatics to obtain  explicit polar optical dispersion relations. In particular, we obtained an LO phonon dispersion relation which is identical to the expression of Sohier {\it et al}. \cite{Sohier:2017}.  
We also derive a 2D longitudinal lattice dielectric function, and further a 2D LST relation for in-plane motion together with a frequency--susceptibility relation for out-of-plane lattice motion. Further we use   
 first-principles calculated quantities  
to study the lattice dynamical properties and also local field and polarizable ion effects in ML TMDs.

This paper is organized as follows. In Section II, a deduction of two pairs of three lattice equations for in-plane and out-of-plane motion is made from a microscopic dipole lattice model. Then the dispersion relations of all the optical branches are obtained together with the phonon group velocity and density of states.  Further the lattice dielectric susceptibility and longitudinal lattice dielectric function are derived and discussed.   
In Section III, we present numerical results of the in-plane optical vibrations in ML TMDs. Using the first-principles calculated parameters of Ref.\cite{Sohier:2017},  
the various lattice dynamical properties including the Born charge and the static and electronic susceptibilities are calculated and compared for several ML TMDs, and further the EP and LFEs are studied in detail.  Finally, Section V summarizes the main results obtained. 


\section{Theory}

\subsection{Equations of motion and lattice polarization}

In this section we use ML MoS$_2$ as representative to present our theory, but the results are applicable to all ML TMDs. Monolayer MoS$_2$ is a 2D binary crystal (with D$_{3h}$ symmetry) with a hexagonal lattice structure (Fig.~\ref{fig1}), which is    
composed of one sublattice of Mo and two sublattices of S, labeled with the base index $\nu=1, 2, \bar{2}$, respectively. Let $a$ and $d$ be the lattice constant and the vertical separation between the two S layers, $d\approx a$ \cite{Ataca:2011,Cao:2012,Chang:2013}, and a lattice cell has an area $s=\sqrt{3}a^2/2$. 
Let $m_{\nu}$ and $e_{\nu}$ be the mass and charge of the type $\nu$ ions, and let $e_2=e_{\bar{2}}$ (i.e., for the two S layers), $e_1=-2e_2=2e_a$, where $2e_a$ is the static effective charge \cite{Karch:1997} on a Mo atom due to the electron charge transfer. 
Let $\boldsymbol{\rho}_l$ be the 2D lattice vector, $\boldsymbol{\rho}_l=l_1\mathbf{a}_1+l_2\mathbf{a}_2$,  $l$ being the cell index, $l=(l_1,l_2)$, and 
let $\mathbf{R}_{l\nu}$ be the position vector of the $\nu$ ion in the $l$-cell, $\mathbf{R}_{l\nu}=\boldsymbol{\rho}_l+\mathbf{r}_{\nu}$, where $\mathbf{r}_{\nu}=(\boldsymbol{\rho}_{\nu},z_{\nu})$, with $z_1=0$, $z_2=d/2$, $z_{\bar{2}}=-d/2$.

In the dipole lattice model \cite{Born:1954} each ion site of type $\nu$ is occupied by an electric dipole $\mathbf{p}_{\nu}$ that is due partly to the ionic displacement $\mathbf{u}_{\nu}$ and partly to the induced electric moment $\boldsymbol{\mu}_{\nu}$ on the ion. 
There is polarization $\mathbf{P}_l=\sum_{\nu}(\mathbf{p}_{\nu}/s)e^{i\mathbf{k}\cdot\mathbf{R}_{l\nu}}\delta(z-z_{\nu})$ associated with a lattice wave of 2D wavevector $\mathbf{k}$. For a {\it long} wavelength lattice wave, $\mathbf{k}\cdot\mathbf{R}_{l\nu}\approx\mathbf{k}\cdot\boldsymbol{\rho}_l$, then 
the lattice can be treated as a polarized continuum with the polarization $\mathbf{P}_c=[\sum_{\nu}\mathbf{p}_{\nu}\delta(z-z_{\nu})]e^{i\mathbf{k}\cdot\boldsymbol{\rho}}/s$, where the discrete lattice vector $\boldsymbol{\rho}_l$ has become a continuous variable $\boldsymbol {\rho}=(x,y)$. Here the $\delta$ function $\delta(z-z_{\nu})$ describes the {\it microscopic} distribution (i.e., at the atomic scale) in the vertical direction of the dipole moment density of the $\nu$ sublattice -- such a microscopic distribution exists as there is no translational symmetry in the $z$ direction. 
When conditions are uniform over many unit cells (for small $k$) the individual layers' specific positions $z_{\nu}$ become unimportant. 
To have a {\it macroscopic} quantity to effectively describe the 2D lattice polarization, therefore, we neglect the difference between the three {\it microscopic} $\delta(z-z_{\nu})$ distributions associated with the three layers of dipoles, by approximating the three $\delta$ functions with a single function independent of the base index $\nu$ which should also capture the {\it microscopic} character of  the polarization's $z$-distribution. We take the function to be an average of the three $\delta$ functions, $\bar{\delta}(z)=\sum_{\nu}\delta(z-z_{\nu})/3$, symmetric with respect to the central Mo layer due to symmetry of the 2D crystal and normalized to unity, such that the 3D polarization of the dielectric is given by $\mathbf{P}=(\sum_{\nu}\mathbf{p}_{\nu}/s)e^{i\mathbf{k}\cdot\boldsymbol{\rho}}\bar{\delta}(z)$. Evidently $\boldsymbol{\mathcal{P}}(\boldsymbol{\rho})=(\sum_{\nu}\mathbf{p}_{\nu}/s)e^{i\mathbf{k}\cdot\boldsymbol{\rho}}$ is a {\it macroscopic} quantity \cite{Born:1954} ({\it average} dipole moment per unit area of the cell, with $\boldsymbol{\mathcal{P}}_0=\sum_{\nu}\mathbf{p}_{\nu}/s$ hereafter) for 2D polarization whilst $\bar{\delta}(z)$ characterizes the volume polarization $\mathbf{P}$' {\it microscopic} distribution in the $z$ direction.  
Physically, the averaging $\bar{\delta}(z)$ plays a role of projecting 
the upper and lower layers of S atoms on the central Mo layer, thus producing a similar hexagonal lattice to ML hBN [Fig.~\ref{fig1}(a)]; this is an effective treatment for the dielectric polarization of the three-layer 2D crystals, which will become more apparent when we obtain the macroscopic fields, lattice modes and dielectric function below.    

We first consider in-plane motion, with the displacements $\mathbf{u}_{\nu}$ and dipole moments $\mathbf{p}_{\nu}$ parallel to  the ML.  The equation of electrostatics is $\nabla\cdot(\mathbf{E}+4\pi\mathbf{P})=0$, where $\mathbf{E}$ is an irrotational macroscopic field, $\mathbf{E}=-\nabla\phi$, $\phi$ being the electric potential, $\phi(\boldsymbol{\rho},z)=\varphi(z)e^{i\mathbf{k}\cdot\boldsymbol{\rho}}$. To solve Poisson's equation 
$\nabla^2\phi(\boldsymbol{\rho},z)=4\pi i\boldsymbol{\mathcal{P}}(\boldsymbol{\rho})\cdot \mathbf{k}\bar{\delta}(z)$, one uses the expansions,  
\begin{equation}
\varphi(z)=\int_{-\infty}^{\infty}\hat{\varphi}(q)e^{iqz}dq,
\label{vphx}
\end{equation}
\begin{equation}
\delta(z-z_{\nu})=\frac{1}{2\pi}\int_{-\infty}^{\infty}e^{iq(z-z_{\nu})}dq. 
\label{delx}
\end{equation}
One finds $\hat{\varphi}(q)=-2i\boldsymbol{\mathcal{P}}_0\cdot \mathbf{k}/[3(k^2+q^2)][1+2\cos(qd/2)]$ and from Eq.~(\ref{vphx}) $\varphi(z)=-2\pi i\boldsymbol{\mathcal{P}}_0\cdot\mathbf{k}\sum_{\nu}e^{-k\lvert z-z_{\nu}\rvert}/(3k)$, and  
 then obtains the parallel and perpendicular components of the macroscopic field:     
\begin{subequations} 
\begin{equation}
\mathbf{E}_{\boldsymbol{\rho}}(\boldsymbol{\rho},z)=-\frac{2\pi}{3}\frac{\mathbf{k}}{k}~\boldsymbol{\mathcal{P}}(\boldsymbol{\rho})\cdot\mathbf{k}\sum_{\nu}e^{-k\lvert z-z_{\nu}\rvert}, 
\label{Ero1}
\end{equation}
\begin{equation}
\mathbf{E}_z(\boldsymbol{\rho},z)=-\mathbf{e}_z\frac{2\pi i}{3}\boldsymbol{\mathcal{P}}(\boldsymbol{\rho})\cdot\mathbf{k}~\sum_{\nu}\sgn(z-z_{\nu})e^{-k\lvert z-z_{\nu}\rvert}.  
\label{Ez1}
\end{equation}
\end{subequations} 

To contrast, when $\mathbf{P}_c$ is used for the polarization, the electric field is given by 
\begin{subequations} 
\begin{equation}
\mathbf{E}_{\boldsymbol{\rho}}(\boldsymbol{\rho},z)=-\frac{2\pi}{s}\frac{\mathbf{k}}{k}~\Big[\sum_{\nu}\mathbf{p}_{\nu}\cdot\mathbf{k}e^{-k\lvert z-z_{\nu}\rvert}\Big]e^{i\mathbf{k}\cdot\boldsymbol{\rho}}, 
\label{Ero1pc}
\end{equation}
\begin{equation}
\mathbf{E}_z(\boldsymbol{\rho},z)=-\mathbf{e}_z\frac{2\pi i}{s}\Big[\sum_{\nu}\mathbf{p}_{\nu}\cdot\mathbf{k}\sgn(z-z_{\nu})e^{-k\lvert z-z_{\nu}\rvert}\Big]e^{i\mathbf{k}\cdot\boldsymbol{\rho}}.  
\label{Ez1pc}
\end{equation}
\end{subequations} 

Comparing Eqs.~(\ref{Ero1}) and (\ref{Ero1pc}), and also Eqs.~(\ref{Ez1}) and (\ref{Ez1pc}) we see that the use of $\bar{\delta}(z)$, i.e., the averaging of the $\delta$ functions leads to the $z$-dependences of the field being substituted by the averages, that is, $e^{-k\lvert z-z_{\nu}\rvert}$ of $\mathbf{E}_{\boldsymbol{\rho}}(\mathbf{r})$ in Eq.~(\ref{Ero1pc}) is replaced by $\sum_{\nu}e^{-k\lvert z-z_{\nu}\rvert}/3$ of Eq.~(\ref{Ero1}), while $\sgn(z-z_{\nu})e^{-k\lvert z-z_{\nu}\rvert}$ of $\mathbf{E}_z(\mathbf{r})$ in Eq.~(\ref{Ez1pc}) is substituted by $\sum_{\nu}\sgn(z-z_{\nu})e^{-k\lvert z-z_{\nu}\rvert}/3$ of Eq.~(\ref{Ez1}), as is expected according to the principle of superposition of electric fields.  
As a result, 
the macroscopic field in the form of Eqs.~(\ref{Ero1}) and (\ref{Ez1}) is {\it proportional} to the 2D macroscopic polarization $\boldsymbol{\mathcal{P}}(\boldsymbol{\rho})$ defined above, as is needed, and further it has captured the limiting behaviour of the electric field of Eqs.~(\ref{Ero1pc}) and (\ref{Ez1pc}), i.e., $\mathbf{E}\rightarrow 0$, as $\mathbf{k}\rightarrow 0$. As $k\lvert z-z_{\nu}\rvert$ is very small at the long wavelengths,  $k\lvert z-z_{\nu}\rvert\leq kd\ll 1$, each $e^{-k\lvert z-z_{\nu}\rvert}$ ($\nu=1, 2, \bar{2}$) varies very slowly in the ML and therefore the averaging is a good treatment for the in-plane field $\mathbf{E}_{\boldsymbol{\rho}}(\mathbf{r})$. The out-of-plane field $\mathbf{E}_z(\mathbf{r})$ of Eq.~(\ref{Ez1}) is zero at $z=0$, and antisymmetric with respect to the central layer on which the upper and lower layers of ions are projected through the averaging $\bar{\delta}(z)$, and thus $\mathbf{E}_z(\mathbf{r})$ has no influence on the ionic motion. Nevertheless, 
only the in-plane field $\mathbf{E}_{\boldsymbol{\rho}}(\mathbf{r})$ is useful, used by the lattice equations, because it is this field that drives the lattice motion. 
Therefore the {\it long} wavelength macroscopic field $\mathbf{E}_{\boldsymbol{\rho}}(\mathbf{r})$ can be considered {\it uniform} within a lattice cell, having the {\it same} value on all the ions in it. 
Here the value of the field is taken at the center of the Mo layer, $z=0$  \cite{Cudazzo:2011}, and we use  
$\mathbf{E}=\mathbf{E}_{\boldsymbol{\rho}}(\boldsymbol{\rho},0)=\mathbf{E}(\boldsymbol{\rho},0)$ for short.

The Lorentz local field $\mathbf{E}_{l}$ (also called the exciting field \cite{Born:1954}), which is the field acting on an ion due to all the {\it other} dipoles in a dipole lattice, is given by the sum of the macroscopic field $\mathbf{E}$ and {\it inner} field  $\mathbf{E}_{in}$ \cite{Born:1954},  
$\mathbf{E}_{l}=\mathbf{E}+\mathbf{E}_{in}$. 
In the {\it long} wavelength lattice waves, the local fields at the Mo and S ion sites are given by the following expressions \cite{Mikhailov:2013}, 
\begin{subequations} 
\begin{equation}
\mathbf{E}_{l,1}=\mathbf{E}+Q_0~\mathbf{p}_1+Q_1~\mathbf{p}_2+Q_1~\mathbf{p}_{\bar{2}}, 
\label{Ex1}
\end{equation}
\begin{equation}
\mathbf{E}_{l,2}=\mathbf{E}+Q_1~\mathbf{p}_1+Q_0~\mathbf{p}_2+Q_2~\mathbf{p}_{\bar{2}}, 
\label{Ex2}
\end{equation}
\begin{equation}
\mathbf{E}_{l,\bar{2}}=\mathbf{E}+Q_1~\mathbf{p}_1+Q_2~\mathbf{p}_2+Q_0~\mathbf{p}_{\bar{2}}. 
\label{Ex2p}
\end{equation}
\end{subequations} 
with the coefficients $Q_0$, $Q_1$ and $Q_2$   
\begin{subequations} 
\begin{equation}
Q_0=\sum_{(m,n)\neq (0,0)}\frac{1}{2(m^2+n^2+mn)^{3/2}a^3}\approx \frac{5.5171}{a^3}~, 
\label{Q0}
\end{equation}
\begin{align}
Q_1&=\sum_{m,n}\Big \{-\frac{1}{[m^2+n^2+mn+n+1/3+(d/a)^2/4]^{3/2}} \Big. 
\nonumber \\ 
&\qquad {}  \Big. +\frac{3(m^2+n^2+mn+n+1/3)}{2[m^2+n^2+mn+n+1/3+(d/a)^2/4]^{5/2}} \Big \}\frac{1}{a^3}, 
\label{Q1}
\end{align}
\begin{align}
Q_2&=\sum_{m,n}\Big \{-\frac{1}{[m^2+n^2+mn+(d/a)^2]^{3/2}} \Big. 
\nonumber \\  &\qquad {} 
 \Big. +\frac{3(m^2+n^2+mn)}{2[m^2+n^2+mn+(d/a)^2]^{5/2}} \Big \}\frac{1}{a^3}. 
\label{Q2}
\end{align}
\end{subequations} 
Note that these $Q$-coefficients and also inner fields are calculated rigorously using the real three-layer crystal structure. 
With the approximation $d=a$, $Q_1$ and $Q_2$ are simplified to $Q_1\approx 1.7184/a^3$ and $Q_2\approx -0.1127/a^3$. 

Eqs.~(\ref{Ex1}), (\ref{Ex2}) and (\ref{Ex2p}) are 2D Lorentz relations for ML MoS$_2$ ($\mathbf{E}$ is the macroscopic field and not an external field in a simple sense \cite{Mikhailov:2013}).   
These relations are valid only for the long lattice waves as in bulk but they have a more complicated form than the 3D Lorentz relation \cite{Born:1954,Kittel:2004} because of the different $Q_i$ ($i=0, 1, 2$) coefficients.

To find the Coulomb field at a $\nu$ ion we also need to consider the field change at the center of the ion due to its own displacement $\mathbf{u}_{\nu}$ \cite{Born:1954}. Evidently the field is equal to the field created at the $\nu$ ion site by displacing all other ions by $-\mathbf{u}_{\nu}$, and therefore equal to the local  field at that ion site in a dipole lattice with displacement dipoles $\mathbf{p}_{\nu'}=-e_{\nu'}\mathbf{u}_{\nu}$, where $\nu'=1, 2, \bar{2}$. Inserting this dipole expression into Eqs.~(\ref{Ex1}), (\ref{Ex2}) and (\ref{Ex2p}) and putting $\mathbf{E}=0$, one finds the field changes at the centers of the Mo and S ions due to their  displacements,  
\begin{subequations} 
\begin{equation}
\mathbf{E}_{u,1}=-\mathbf{u}_1(e_1Q_0+e_2Q_1+e_{\bar{2}}Q_1), 
\label{Eu1}
\end{equation}
\begin{equation}
\mathbf{E}_{u,2}=-\mathbf{u}_2(e_1Q_1+e_2Q_0+e_{\bar{2}}Q_2),  
\label{Eu2}
\end{equation}
\begin{equation}
\mathbf{E}_{u,\bar{2}}=-\mathbf{u}_{\bar{2}}(e_1Q_1+e_2Q_2+e_{\bar{2}}Q_0).   
\label{Eu2p}
\end{equation}
\end{subequations} 
The {\it total} Coulomb fields $\mathbf{E}_1$, $\mathbf{E}_2$ and $\mathbf{E}_{\bar{2}}$ at the centers of the Mo and S ions are  the sums of $\mathbf{E}_{l,1}$ and $\mathbf{E}_{u,1}$  [Eqs.~(\ref{Ex1}) and (\ref{Eu1})],  $\mathbf{E}_{l,2}$ and $\mathbf{E}_{u,2}$ [Eqs.~(\ref{Ex2}) and (\ref{Eu2})], and $\mathbf{E}_{l,\bar{2}}$ and $\mathbf{E}_{u,\bar{2}}$ [Eqs.~(\ref{Ex2p}) and (\ref{Eu2p})], respectively, 
\begin{subequations} 
\begin{equation}
\mathbf{E}_1=\mathbf{E}+Q_0~\mathbf{p}_1+Q_1~\mathbf{p}_2+Q_1~\mathbf{p}_{\bar{2}}+2e_a(Q_1-Q_0)\mathbf{u}_1,  
\label{Eto1}
\end{equation}
\begin{equation}
\mathbf{E}_2=\mathbf{E}+Q_1~\mathbf{p}_1+Q_0~\mathbf{p}_2+Q_2~\mathbf{p}_{\bar{2}}+e_a(Q_0+Q_2-2Q_1)\mathbf{u}_2,  
\label{Eto2}
\end{equation}
\begin{equation}
\mathbf{E}_{\bar{2}}=\mathbf{E}+Q_1~\mathbf{p}_1+Q_2~\mathbf{p}_2+Q_0~\mathbf{p}_{\bar{2}} +e_a(Q_0+Q_2-2Q_1)\mathbf{u}_{\bar{2}},  
\label{Eto2p}
\end{equation}
\end{subequations} 
where all the vectors are parallel to the ML. 

The {\it electronic} polarization of an ion is equivalent to a point-dipole so the {\it induced} dipole moment of a $\nu$ ion is $\boldsymbol{\mu}_{\nu}=\alpha_{\nu}\mathbf{E}_{\nu}$ \cite{Born:1954}, where $\alpha_{\nu}$ is the in-plane {\it electronic} polarizability of the $\nu$ ions ($\alpha_2=\alpha_{\bar{2}}$).  
Then the {\it total} dipole moments on the Mo and S ions are 
\begin{subequations} 
\begin{equation}
\mathbf{p}_1=2e_a\mathbf{u}_1+\alpha_1~\mathbf{E}_1, 
\label{pto1}
\end{equation}
\begin{equation}
\mathbf{p}_2=-e_a\mathbf{u}_2+\alpha_2~\mathbf{E}_2,  
\label{pto2}
\end{equation}
\begin{equation}
\mathbf{p}_{\bar{2}}=-e_a\mathbf{u}_{\bar{2}}+\alpha_2~\mathbf{E}_{\bar{2}}. 
\label{pto2p}
\end{equation}
\end{subequations} 
When the expressions for the total fields $\mathbf{E}_1$, $\mathbf{E}_2$ and $\mathbf{E}_{\bar{2}}$ are inserted, one finds  
\begin{subequations} 
\begin{align}
&(1-\alpha_1Q_0)\mathbf{p}_1-\alpha_1Q_1(\mathbf{p}_2+\mathbf{p}_{\bar{2}})
\nonumber \\ &\qquad {} 
 =2e_a[1+\alpha_1(Q_1-Q_0)]\mathbf{u}_1+\alpha_1~\mathbf{E}, 
\label{pto1a}
\end{align}
\begin{align}
&-\alpha_2Q_1\mathbf{p}_1+(1-\alpha_2Q_0)\mathbf{p}_2-\alpha_2Q_2\mathbf{p}_{\bar{2}}
\nonumber \\ &\qquad {} 
=-e_a[1-\alpha_2(Q_0+Q_2-2Q_1)]\mathbf{u}_2+\alpha_2~\mathbf{E},  
\label{pto2a}
\end{align}
\begin{align}
&-\alpha_2Q_1\mathbf{p}_1-\alpha_2Q_2\mathbf{p}_2+(1-\alpha_2Q_0)\mathbf{p}_{\bar{2}}
\nonumber \\ &\qquad {} 
=-e_a[1-\alpha_2(Q_0+Q_2-2Q_1)]\mathbf{u}_{\bar{2}}+\alpha_2~\mathbf{E}.   
\label{pto2pa}
\end{align}
\end{subequations} 
Solving these equations, one can express $\mathbf{p}_1$, $\mathbf{p}_2$ and $\mathbf{p}_{\bar{2}}$ in terms of $\mathbf{u}_1$, $\mathbf{u}_2$, $\mathbf{u}_{\bar{2}}$ and $\mathbf{E}$ as follows: 
\begin{subequations} 
\begin{align}
\mathbf{p}_1&=\frac{1}{D}\Big \{2e_a\big[1-\alpha_2(Q_0+Q_2)\big]\big[1+\alpha_1(Q_1-Q_0)\big]\mathbf{u}_1 \Big.
\nonumber \\ &\qquad {} -2e_a\alpha_1Q_1\big[1-\alpha_2(Q_0+Q_2-2Q_1)\big]\mathbf{u}_c  
\nonumber \\ &\qquad {}  \Big. +\alpha_1\big[1-\alpha_2(Q_0+Q_2-2Q_1)\big]\mathbf{E} \Big \}, 
\label{pt1}
\end{align}
\begin{align}
\mathbf{p}_2&=\frac{1}{D}\Big \{2e_a\alpha_2Q_1\big[1+\alpha_1(Q_1-Q_0)\big]\mathbf{u}_1 \Big.
\nonumber \\ &\qquad {} -e_a(1-\alpha_1Q_0)\big[1-\alpha_2(Q_0+Q_2-2Q_1)\big]\mathbf{u}_c   
\nonumber \\ &\qquad {}  \Big. +\alpha_2\big[1+\alpha_1(Q_1-Q_0)\big]\mathbf{E} \Big \}
-\frac{1}{2}e_a\eta\mathbf{u}_d,    
\label{pt2}
\end{align}
\begin{align}
\mathbf{p}_{\bar{2}}&=\frac{1}{D}\Big \{2e_a\alpha_2Q_1\big[1+\alpha_1(Q_1-Q_0)\big]\mathbf{u}_1 \Big. 
\nonumber \\ &\qquad {} -e_a(1-\alpha_1Q_0)\big[1-\alpha_2(Q_0+Q_2-2Q_1)\big]\mathbf{u}_c  
\nonumber \\ &\qquad {}  \Big. +\alpha_2\big[1+\alpha_1(Q_1-Q_0)\big]\mathbf{E} \Big \}
+\frac{1}{2}e_a\eta\mathbf{u}_d,    
\label{pt2p}
\end{align}
\end{subequations} 
where  $\mathbf{u}_d$ and $\mathbf{u}_c$ are the relative displacement of the two S atoms and the displacement of their centre of mass, respectively,    
\begin{subequations} 
\begin{equation}
\mathbf{u}_d=\mathbf{u}_2-\mathbf{u}_{\bar{2}},  
\label{udss}
\end{equation}
\begin{equation}
\mathbf{u}_c=(\mathbf{u}_2+\mathbf{u}_{\bar{2}})/2,  
\label{udss}
\end{equation}
\end{subequations} 
and 
\begin{equation}
D=(1-\alpha_1Q_0)[1-\alpha_2(Q_0+Q_2)]-2\alpha_1\alpha_2Q_1^2,   
\label{bigD}
\end{equation}
\begin{equation}
\eta=\frac{1-\alpha_2(Q_0+Q_2-2Q_1)}{1-\alpha_2(Q_0-Q_2)}.  
\label{eta3p}
\end{equation}

Introducing the optical displacement $\mathbf{w}$ to describe the motion of the Mo atom relative to the two S atoms, 
\begin{equation}
\mathbf{w}=\sqrt{\frac{\bar{m}}{s}}(\mathbf{u}_1-\mathbf{u}_c)=\sqrt{\frac{\bar{m}}{s}}\left[\mathbf{u}_1-\frac{1}{2}(\mathbf{u}_2+\mathbf{u}_{\bar{2}})\right],    
\label{smw}
\end{equation}
with $\bar{m}$ being the reduced mass, $\bar{m}=2m_1m_2/(m_1+2m_2)$,  one finds that 
the {\it areal} polarization $\boldsymbol{\mathcal{P}}=(\mathbf{p}_1+\mathbf{p}_2+\mathbf{p}_{\bar{2}})/s$  is given by 
\begin{equation}
\boldsymbol{\mathcal{P}}=a_{21}\mathbf{w}+a_{22}\mathbf{E},   
\label{bigP1}
\end{equation}
where 
\begin{subequations} 
\begin{equation}
a_{21}=\frac{2e_a}{D\sqrt{\bar{m}s}}\big[1+\alpha_1(Q_1-Q_0)\big]\big[1-\alpha_2(Q_0+Q_2-2Q_1)\big], 
\label{a21}
\end{equation}
\begin{align}
a_{22}&=\frac{1}{sD}\left\{\alpha_1\big[1-\alpha_2(Q_0+Q_2-2Q_1)\big] \right. 
\nonumber \\ &\qquad {} \left. +2\alpha_2\big[1+\alpha_1(Q_1-Q_0)\big]\right\}.   
\label{a22}
\end{align}
\end{subequations} 
Eq.~(\ref{bigP1}) shows that the 2D {\it macroscopic} polarization $\boldsymbol{\mathcal{P}}$ of ML MoS$_2$ reduces to a sum of two contributions, one due to the optical displacement and the other due to the {\it macroscopic} field. 

It is evident from Eq.~(\ref{bigP1}) that $a_{22}$ is the in-plane {\it electronic} susceptibility $\chi_e$ of 2D MoS$_2$, 
\begin{equation}
a_{22}=\chi_e.      
\label{a22alf}
\end{equation} 
Expression (\ref{a22}) relates the dielectric susceptibility to the  polarizabilities of the constituent atoms, and is a Clausius-Mossotti relation. 
Introducing the Born charge \cite{Gonze:1997} 
\begin{equation}
e_B=\frac{2e_a}{D}\big[1+\alpha_1(Q_1-Q_0)\big]\big[1-\alpha_2(Q_0+Q_2-2Q_1)\big], 
\label{eB}
\end{equation}
then the coefficient $a_{21}$ relates to $e_B$ through  
\begin{equation}
a_{21}=\frac{e_B}{\sqrt{\bar{m}s}}.   
\label{a21b}
\end{equation}

Using Eqs.~(\ref{bigD}), (\ref{a22}), (\ref{a22alf}) and (\ref{eB}) one obtains   
\begin{equation}
a_{22}=\chi_e=\frac{1}{sQ_1}\left(\frac{e_B}{2e_a}-1\right), 
\label{a22eB}
\end{equation}
showing that the $a_{22}$ coefficient is also related to the Born charge. $e_B \neq 2e_a$ owing to the EP of ions, i.e., $\chi_e>0$,  and as $Q_1 >0$, $e_B$ and $e_a$ have the same sign and $\lvert e_B\rvert >2\lvert e_a\rvert$.  

Insert expressions (\ref{pt1}), (\ref{pt2}) and (\ref{pt2p}) for $\mathbf{p}_1$ and $\mathbf{p}_2$  and $\mathbf{p}_{\bar{2}}$ into Eqs.~(\ref{Eto1}), (\ref{Eto2}) and (\ref{Eto2p}) and express the total fields $\mathbf{E}_1$, $\mathbf{E}_2$ and $\mathbf{E}_{\bar{2}}$ in terms of the ionic displacements and macroscopic field. After a lengthy algebraic simplification the total field expressions are simplified to
\begin{subequations} 
\begin{equation}
\mathbf{E}_1=\frac{1}{D}\big[1-\alpha_2(Q_0+Q_2-2Q_1)\big]\big[2e_aQ_1(\mathbf{u}_1-\mathbf{u}_c)+\mathbf{E}\big], 
\label{Eto1b}
\end{equation}
\begin{align}
\mathbf{E}_2&=\frac{2e_a}{D}\Big \{Q_1\big[1+\alpha_1(Q_1-Q_0)\big]\mathbf{u}_1 \Big . 
\nonumber \\ &\qquad {} -\big[1-\alpha_2(Q_0+Q_2-2Q_1)\big]
\nonumber \\ &\qquad {} \Big . 
\big[\alpha_1Q_1^2+\frac{1}{2}(1-\alpha_1Q_0)(Q_0+Q_2)\big]\mathbf{u}_c  \Big \}  
\nonumber \\ 
&\qquad {} -\frac{1}{2}(Q_0-Q_2)e_a\eta\mathbf{u}_d+e_a(Q_0+Q_2-2Q_1)\mathbf{u}_2
\nonumber \\ &\qquad {} +\frac{1}{D}\big[1+\alpha_1(Q_1-Q_0)\big]\mathbf{E},  
\label{Eto2b}
\end{align}
\begin{align}
\mathbf{E}_{\bar{2}}&=\frac{2e_a}{D}\Big \{Q_1\big[1+\alpha_1(Q_1-Q_0)\big]\mathbf{u}_1 \Big .
\nonumber \\ &\qquad {} -\big[1-\alpha_2(Q_0+Q_2-2Q_1)\big]
\nonumber \\ &\qquad {} \Big . \big[\alpha_1Q_1^2+\frac{1}{2}(1-\alpha_1Q_0)(Q_0+Q_2)\big]\mathbf{u}_c  \Big \}  
\nonumber \\ 
&\qquad {} +\frac{1}{2}(Q_0-Q_2)e_a\eta\mathbf{u}_d+e_a(Q_0+Q_2-2Q_1)\mathbf{u}_{\bar{2}}
\nonumber \\ &\qquad {} +\frac{1}{D}\big[1+\alpha_1(Q_1-Q_0)\big]\mathbf{E}.  
\label{Eto2pb}
\end{align}
\end{subequations} 
Further one readily finds from Eqs.~(\ref{Eto2b}), (\ref{Eto2pb}) 
\begin{subequations} 
\begin{equation}
\mathbf{E}_2+\mathbf{E}_{\bar{2}}=\frac{2}{D}\big[1+\alpha_1(Q_1-Q_0)\big]\big[2e_aQ_1(\mathbf{u}_1-\mathbf{u}_c)+\mathbf{E}\big], 
\label{Eto22pb}
\end{equation}
\begin{equation}
\mathbf{E}_2-\mathbf{E}_{\bar{2}}=-\frac{2e_a(Q_1-Q_2)}{1-\alpha_2(Q_0-Q_2)}\mathbf{u}_d.  
\label{Eto22mb}
\end{equation}
\end{subequations}

The Coulomb force on a $\nu$ ion consists of two parts \cite{Born:1954},  (i) the force exerted on ionic charge $e_{\nu}$ by total field $\mathbf{E}_{\nu}$, and (ii) the force exerted on {\it induced} dipole $\boldsymbol{\mu}_{\nu}$ by the field of all other ions. 
The latter can be obtained by using again the Lorentz relations  (\ref{Ex1}), (\ref{Ex2}) and (\ref{Ex2p}). Subject dipole $\boldsymbol{\mu}_{\nu}$ at a particular site of ion $\nu$ a virtual displacement $\mathbf{u}$ whilst fixing all other ions in their equilibrium positions. The virtual energy is the interaction energy between the dipole and the {\it field} created equivalently at the ion site by displacing all other ions by $-\mathbf{u}$ \cite{Born:1954}, which is exactly the {\it local field} in the  dipole lattice with dipole moments $\mathbf{p}_{\nu^\prime}=-e_{\nu^\prime}\mathbf{u}$, ($\nu^\prime=1, 2, \bar{2}$). Therefore the virtual energy is 
$-\boldsymbol{\mu}_{\nu}\cdot\mathbf{E}_{l,\nu}$, resulting in the force on the dipole $\nabla _{\mathbf{u}}(\boldsymbol{\mu}_{\nu}\cdot\mathbf{E}_{l,\nu})$. Inserting the dipole moment above into Eqs.~(\ref{Ex1}), (\ref{Ex2})  and (\ref{Ex2p}) and putting $\mathbf{E}=0$, one obtains the local fields and then the forces on the dipoles $\boldsymbol{\mu}_1$, $\boldsymbol{\mu}_2$ and $\boldsymbol{\mu}_{\bar{2}}$ given by   
$2e_a(Q_1-Q_0)\boldsymbol{\mu}_1$, $e_a(Q_0+Q_2-2Q_1)\boldsymbol{\mu}_2$, and $e_a(Q_0+Q_2-2Q_1)\boldsymbol{\mu}_{\bar{2}}$, respectively, where $\boldsymbol{\mu}_{\nu}=\alpha_{\nu}\mathbf{E}_{\nu}$ ($\nu=1, 2, \bar{2}$).   

To describe the restoring forces, 
let $K_1$ be the spring force constant between the Mo and S ions, and $K_2$ be the spring force constant between the two S layers. 
Therefore the equations of motion for the Mo and S ions are given by 
\begin{subequations} 
\begin{align}
m_1\ddot{\mathbf{u}}_1&=-K_1(\mathbf{u}_1-\mathbf{u}_2)-K_1(\mathbf{u}_1-\mathbf{u}_{\bar{2}})
\nonumber \\ &\qquad {} +2e_a[1+\alpha_1(Q_1-Q_0)]\mathbf{E}_1, 
\label{eom1}
\end{align}
\begin{align}
m_2\ddot{\mathbf{u}}_2&=-K_1(\mathbf{u}_2-\mathbf{u}_1)-K_2(\mathbf{u}_2-\mathbf{u}_{\bar{2}})
\nonumber \\ &\qquad {} -e_a[1-\alpha_2(Q_0+Q_2-2Q_1)]\mathbf{E}_2,   
\label{eom2}
\end{align}
\begin{align}
m_2\ddot{\mathbf{u}}_{\bar{2}}&=-K_1(\mathbf{u}_{\bar{2}}-\mathbf{u}_1)+K_2(\mathbf{u}_2-\mathbf{u}_{\bar{2}})
\nonumber \\ &\qquad {} -e_a[1-\alpha_2(Q_0+Q_2-2Q_1)]\mathbf{E}_{\bar{2}}.    
\label{eom2p}
\end{align}
\end{subequations} 

Substitute expression (\ref{Eto1b}) for $\mathbf{E}_1$ in Eq.~(\ref{eom1}). Add Eqs.~(\ref{eom2}) and (\ref{eom2p}), change  $\mathbf{u}_2+\mathbf{u}_{\bar{2}}$ to $2\mathbf{u}_c$, and substitute expression (\ref{Eto22pb}) for $\mathbf{E}_2+\mathbf{E}_{\bar{2}}$. Subtract Eqs.~(\ref{eom2}) and (\ref{eom2p}), change  $\mathbf{u}_2-\mathbf{u}_{\bar{2}}$ to $\mathbf{u}_d$, and then insert expression (\ref{Eto22mb}) for $\mathbf{E}_2-\mathbf{E}_{\bar{2}}$. Introducing a force constant due to LFEs, $K_e$, for the polar optical motion,   
\begin{equation}
K_e=e_ae_BQ_1, 
\label{Ke}
\end{equation}
and another force constant due to LFEs, $K_{2,d}$, for the nonpolar motion, 
\begin{equation}
K_{2,d}=e_a^2(Q_1-Q_2)\frac{1-\alpha_2(Q_0+Q_2-2Q_1)}{1-\alpha_2(Q_0-Q_2)}, 
\label{Kd}
\end{equation}
and using the Born charge $e_B$ [expression (\ref{eB})],  the equations of motion reduce to  
\begin{subequations} 
\begin{equation}
m_1\ddot{\mathbf{u}}_1=2(-K_1+K_e)(\mathbf{u}_1-\mathbf{u}_c)+e_B\mathbf{E}, 
\label{eom1b}
\end{equation}
\begin{equation}
2m_2\ddot{\mathbf{u}}_c=2(K_1-K_e)(\mathbf{u}_1-\mathbf{u}_c)-e_B\mathbf{E},   
\label{eom22pb}
\end{equation}
\begin{equation}
\frac{1}{2}m_2\ddot{\mathbf{u}}_d=-\Big(\frac{K_1}{2}+K_2-K_{2,d}\Big)\mathbf{u}_d.  
\label{eom22mb}
\end{equation}
\end{subequations} 

Multiplying Eqs.~(\ref{eom1b}) and (\ref{eom22pb}) by $2m_2$ and $m_1$ respectively, subtracting and then dividing by $(m_1+2m_2)$, one finds 
\begin{equation}
\bar{m}(\ddot{\mathbf{u}}_1-\ddot{\mathbf{u}}_c)=2(K_e-K_1)(\mathbf{u}_1-\mathbf{u}_c)+e_B\mathbf{E}.  
\label{eom12b}
\end{equation}

Changing $\mathbf{u}_1-\mathbf{u}_c$ to $\mathbf{w}$ with Eq.~(\ref{smw}), one obtains  
\begin{equation}
\ddot{\mathbf{w}}=a_{11}\mathbf{w}+a_{12}\mathbf{E},   
\label{eomw1}
\end{equation}
where
\begin{subequations} 
\begin{equation}
a_{11}=\frac{2}{\bar{m}}(K_e-K_1)=-\omega_0^2, 
\label{a11}
\end{equation}
\begin{equation}
a_{12}=\frac{e_B}{\sqrt{\bar{m}s}},
\label{a12}
\end{equation}
\end{subequations} 
$\omega_0$ being the intrinsic oscillator frequency (i.e., without $\mathbf{E}$) for the polar optical vibrations.  

From Eqs.~(\ref{a21b}) and (\ref{a12}) one finds two {\it equal} $a$-coefficients 
\begin{equation}
a_{12}=a_{21}.  
\label{a12a21}
\end{equation}

Eq.~(\ref{eom22mb}) is the equation of motion for the optical vibrations of the two S ions with reduced mass $m_2/2$;  
introducing $\mathbf{w}_d$, 
\begin{equation}
\mathbf{w}_d=\sqrt{\frac{m_2}{2s}}\mathbf{u}_d,   
\label{wdsso1}
\end{equation}
the equation becomes   
\begin{equation}
\ddot{\mathbf{w}}_d=a_d\mathbf{w}_d,   
\label{eomwss1}
\end{equation}
where
\begin{equation}
a_d=-(K_1+2K_2-2K_{2,d})/m_2=-\omega_{2\bar{2}}^2, 
\label{add1}
\end{equation}
$\omega_{2\bar{2}}$ being the frequency of the long-wavelength in-plane nonpolar  modes, i.e., the LO$_1$ and TO$_1$ modes, which are nondispersive, and doubly degenerate at $\Gamma$.   

The equation of motion (\ref{eomw1}) and polarization equation (\ref{bigP1}) describe the  in-plane {\it polar} optical vibrations of  ML MoS$_2$, the vibrational properties of which will be derived in Sec. III below.  As the lattice vibrations considered here are of long wavelengths, the pair of equations (\ref{eomw1}) and (\ref{bigP1}) constitute a macroscopic description of the lattice motion. 
$\mathbf{E}$ appearing in these equations represents the in-plane component of the macroscopic field in the thin ML and the $\mathbf{E}$ at $z=0$ is used (as was detailed above), i.e., $\mathbf{E}=\mathbf{E}_{\boldsymbol{\rho}}(\boldsymbol{\rho},0)$.  
It is evident from Eqs.~(\ref{eom1b}) and (\ref{eom22pb}) that the center of mass of the three ions in the cell remains stationary  (frequency $\omega=0$), yielding trivial nondynamical solutions.  

Now we consider out-of-plane motion with displacements $\mathbf{u}_{\nu}$ and dipole moments $\mathbf{p}_{\nu}$ parallel to $\mathbf{e}_z$. 
Let $\alpha_{\nu}^\prime$ be the electronic polarizability of the $\nu$ ions; usually $\alpha_{\nu}^\prime\neq\alpha_{\nu}$ ($\alpha_{\nu}$ is the in-plane polarizability), both being components of the polarizability tensor \cite{Born:1954}. Let $K_1^\prime$ be the spring force constant between the Mo and S ions and $K_2^\prime$ be the force constant between the two S layers.  Let $2e_a^\prime$ denote the static {\it effective} charge on the Mo atoms, which may differ from the in-plane {\it effective} charge $2e_a$ owing to  the anisotropic 3D charge density distribution \cite{Zhang:2019b};  thus the charges on the ions are $e_1^\prime=2e_a^\prime$, $e_2^\prime=e_{\bar{2}}^\prime=-e_a^\prime$.    
Solving the Poisson equation one finds the electrostatic potential
\begin{equation}
\varphi(z)=\frac{2\pi}{3} \boldsymbol{\mathcal{P}}_0\cdot\mathbf{e}_z\sum_{\nu}\sgn(z-z_{\nu})e^{-k\lvert z-z_{\nu}\rvert}, 
\label{Pophez}
\end{equation}
and then the macroscopic field,     
\begin{subequations} 
\begin{equation}
\mathbf{E}_z(\boldsymbol{\rho},z)=-4\pi\boldsymbol{\mathcal{P}}_0\Big[\bar{\delta}(z)-\frac{1}{6}k\sum_{\nu}e^{-k\lvert z-z_{\nu}\rvert}\Big]e^{i\mathbf{k}\cdot\boldsymbol{\rho}},   
\label{Ez2}
\end{equation}
\begin{equation}
\mathbf{E}_{\boldsymbol{\rho}}(\boldsymbol{\rho},z)=-\frac{2\pi i}{3}\mathbf{k}~\boldsymbol{\mathcal{P}}_0\cdot\mathbf{e}_z\sum_{\nu}\sgn(z-z_{\nu})e^{-k\lvert z-z_{\nu}\rvert}e^{i\mathbf{k}\cdot\boldsymbol{\rho}}.  
\label{Ero2}
\end{equation}
\end{subequations} 
Evidently the in-plane component of the field is zero in all the three atomic layers. 
The local fields (i.e., Lorentz relations) at the Mo and two S sites are given by   
\begin{subequations} 
\begin{equation}
\mathbf{E}_{l,1}=\mathbf{E}+Q_0^\prime~\mathbf{p}_1+Q_1^\prime~\mathbf{p}_2+Q_1^\prime~\mathbf{p}_{\bar{2}}, 
\label{Ex1z}
\end{equation}
\begin{equation}
\mathbf{E}_{l,2}=\mathbf{E}+Q_1^\prime~\mathbf{p}_1+Q_0^\prime~\mathbf{p}_2+Q_2^\prime~\mathbf{p}_{\bar{2}},  
\label{Ex2z}
\end{equation}
\begin{equation}
\mathbf{E}_{l,\bar{2}}=\mathbf{E}+Q_1^\prime~\mathbf{p}_1+Q_2^\prime~\mathbf{p}_2+Q_0^\prime~\mathbf{p}_{\bar{2}},  
\label{Ex2z}
\end{equation}
\end{subequations} 
respectively, where the $Q^\prime$ coefficients are $Q_0^\prime=-2Q_0$, $Q_1^\prime=-2Q_1$ and $Q_2^\prime=-2Q_2$ from a symmetry analysis \cite{Mikhailov:2013}. The macroscopic field has a $\bar{\delta}(z)$ term [Eq.~(\ref{Ez2})] but it makes no contribution to the field change experienced by an ion $\nu$ owing to its own displacement $\mathbf{u}_{\nu}$, i.e., $4\pi\mathbf{u}_{\nu}\sum_{\nu^\prime}e_{\nu^\prime}^\prime\bar{\delta}(z)/s=4\pi\mathbf{u}_{\nu}(e_1^\prime+e_2^\prime+e_{\bar{2}}^\prime)\bar{\delta}(z)/s=0$. 
Thus the field changes at the centers of Mo and S owing to their own displacements are   
\begin{subequations} 
\begin{equation}
\mathbf{E}_{u,1}=-\mathbf{u}_1(e_1^\prime Q_0^\prime+e_2^\prime Q_1^\prime+e_{\bar{2}}^\prime Q_1^\prime), 
\label{Eu1z}
\end{equation}
\begin{equation}
\mathbf{E}_{u,2}=-\mathbf{u}_2(e_1^\prime Q_1^\prime+e_2^\prime Q_0^\prime+e_{\bar{2}}^\prime Q_2^\prime),  
\label{Eu2z}
\end{equation}
\begin{equation}
\mathbf{E}_{u,{\bar{2}}}=-\mathbf{u}_{\bar{2}}(e_1^\prime Q_1^\prime+e_2^\prime Q_2^\prime+e_{\bar{2}}^\prime Q_0^\prime).   
\label{Eu2pz}
\end{equation}
\end{subequations} 
Similarly, the macroscopic field has no contribution to the force exerted on the induced dipole $\boldsymbol{\mu}_{\nu}$ by all other ions again because of the charge neutrality in each cell. 
Repeating the process as before, one finds that with replacements $Q_0\rightarrow Q_0^\prime$, $Q_1\rightarrow Q_1^\prime$, $Q_2\rightarrow Q_2^\prime$, $e_a\rightarrow e_a^\prime$, $\alpha_{\nu}\rightarrow\alpha_{\nu}^\prime$, $ K_1\rightarrow K_1^\prime$ and $ K_2\rightarrow K_2^\prime$, all the equations above for in-plane motion are applicable to the out-of-plane motion. 
The Born charge is 
\begin{equation}
e_B^\prime=\frac{2e_a^\prime}{D^\prime}\big[1+\alpha_1^\prime(Q_1^\prime-Q_0^\prime)\big]\big[1-\alpha_2^\prime(Q_0^\prime+Q_2^\prime-2Q_1^\prime)\big], 
\label{eBz}
\end{equation}
where
\begin{equation}
D^\prime=(1-\alpha_1^\prime Q_0^\prime)[1-\alpha_2^\prime(Q_0^\prime+Q_2^\prime)]-2\alpha_1^\prime\alpha_2^\prime Q_1^{\prime 2}.  
\label{bigD1}
\end{equation}
The polarization equation is 
\begin{equation}
\boldsymbol{\mathcal{P}}=c_{21}\mathbf{w}+c_{22}\mathbf{E},   
\label{bigP1z}
\end{equation}
where $\mathbf{w}$ is the optical displacement [Eq.~(\ref{smw})] and $c_{22}$ is out-of-plane electronic susceptibility $\chi_e^\prime$, 
\begin{equation}
c_{22}=\chi_e^\prime,     
\label{c22alf}
\end{equation} 
and both $c_{21}$ and $c_{22}$ are related to the Born charge,   
\begin{subequations} 
\begin{equation}
c_{21}=\frac{e_B^\prime}{\sqrt{\bar{m}s}},  
\label{c21z}
\end{equation}
\begin{equation}
c_{22}=\chi_e^\prime=\frac{1}{sQ_1^\prime}\left(\frac{e_B^\prime}{2e_a^\prime}-1\right). 
\label{c22z}
\end{equation}
\end{subequations}

The equation of motion is  
\begin{equation}
\ddot{\mathbf{w}}=c_{11}\mathbf{w}+c_{12}\mathbf{E},   
\label{eomw1z}
\end{equation}
where
\begin{subequations} 
\begin{equation}
c_{11}=\frac{2}{\bar{m}}(K_e^\prime-K_1^\prime)=-\omega_0'^2 \quad  (K_e^\prime=e_a^\prime e_B^\prime Q_1^\prime), \label{c11z}
\end{equation}
\begin{equation}
c_{12}=c_{21}=\frac{e_B^\prime}{\sqrt{\bar{m}s}}, 
\label{c12z}
\end{equation}
\end{subequations} 
$\omega_0'$ being the intrinsic oscillator frequency for the polar optical vibrations. 

For the out-of-plane nonpolar optical vibrations the equation of motion is given by  
\begin{equation}
\ddot{\mathbf{w}}_d=c_d\mathbf{w}_d,   
\label{eomwss1z}
\end{equation}
where 
\begin{equation}
\mathbf{w}_d=\sqrt{\frac{m_2}{2s}}\Big(\mathbf{u}_2-\mathbf{u}_{\bar{2}}\Big),   
\label{wdsso1z}
\end{equation}
and 
\begin{equation}
c_d=-(K_1^\prime+2K_2^\prime-2K_{2,d}^\prime)/m_2=-\omega_{2\bar{2}}^{\prime 2}, 
\label{cdd1z}
\end{equation}
with 
\begin{equation}
K_{2,d}^\prime=e_a^{\prime 2}(Q_1^\prime-Q_2^\prime)\frac{1-\alpha_2^\prime(Q_0^\prime+Q_2^\prime-2Q_1^\prime)}{1-\alpha_2^\prime(Q_0^\prime-Q_2^\prime)}, 
\label{Kd1z}
\end{equation}
and $\omega_{2\bar{2}}^\prime$ being the frequency of the long-wavelength nonpolar modes, i.e., the ZO$_1$ modes.

In the lattice equations (\ref{eomw1z}) and (\ref{bigP1z}) $\mathbf{E}$ represents the field in the thin ML taking $\mathbf{E}(\boldsymbol{\rho},0)$, and evidently $\mathbf{E}(\boldsymbol{\rho},0)=\mathbf{E}_z(\boldsymbol{\rho},0)$ for the out-of-plane  vibrations.

When the equations for the in-plane motion [Eqs.~(\ref{eomw1}), (\ref{bigP1}) and (\ref{eomwss1})] and out-of-plane motion [Eqs.~(\ref{eomw1z}), (\ref{bigP1z}) and (\ref{eomwss1z})] are considered simultaneously, they can be rewritten for clarity as 
\begin{subequations} 
\begin{equation}
\ddot{\mathbf{w}}_{\boldsymbol{\rho}}(\boldsymbol{\rho})=a_{11}\mathbf{w}_{\boldsymbol{\rho}}(\boldsymbol{\rho})+a_{12}\mathbf{E}_{\boldsymbol{\rho}}(\boldsymbol{\rho},0),   
\label{eomw1co}
\end{equation}
\begin{equation}
\boldsymbol{\mathcal{P}}_{\boldsymbol{\rho}}(\boldsymbol{\rho})=a_{21}\mathbf{w}_{\boldsymbol{\rho}}(\boldsymbol{\rho})+a_{22}\mathbf{E}_{\boldsymbol{\rho}}(\boldsymbol{\rho},0),   
\label{bigP1co}
\end{equation}
\begin{equation}
\ddot{\mathbf{w}}_{d,\boldsymbol{\rho}}(\boldsymbol{\rho})=a_d\mathbf{w}_{d,\boldsymbol{\rho}}(\boldsymbol{\rho}),   
\label{eomwss1co}
\end{equation}
\end{subequations} 
and 
\begin{subequations} 
\begin{equation}
\ddot{\mathbf{w}}_z(\boldsymbol{\rho})=c_{11}\mathbf{w}_z(\boldsymbol{\rho})+c_{12}\mathbf{E}_z(\boldsymbol{\rho},0),   
\label{eomw1zco}
\end{equation}
\begin{equation}
\boldsymbol{\mathcal{P}}_z(\boldsymbol{\rho})=c_{21}\mathbf{w}_z(\boldsymbol{\rho})+c_{22}\mathbf{E}_z(\boldsymbol{\rho},0),   
\label{bigP1zco}
\end{equation}
\begin{equation}
\ddot{\mathbf{w}}_{d,z}(\boldsymbol{\rho})=c_d\mathbf{w}_{d,z}(\boldsymbol{\rho}),   
\label{eomwss1zco}
\end{equation}
\end{subequations} 
respectively, where $a_{12}=a_{21}$ and $c_{12}=c_{21}$, and the linear coefficients have been given by the expressions above. The two pairs of equations involving  macroscopic field have the same forms as those for ML hBN \cite{Zhang:2019b} and are also similar to  Huang's equations for bulk crystals \cite{Huang:1951,Huang:1951a,Born:1954}, whereas the other two equations show the effective forces in a similar form to the Hooke's law formula.   
By means of the averaging $\bar{\delta}(z)$, not only the 2D macroscopic polarization is introduced but also the 2D lattice equations are obtained with two pairs of equal linear coefficients.

The relation $a_{12}=a_{21}$ allows us to introduce an energy density (energy per unit area) $u_p$ associated with the in-plane optical vibrations, as a function of $\mathbf{w}_{\boldsymbol{\rho}}(\boldsymbol{\rho})$, $\mathbf{E}_{\boldsymbol{\rho}}(\boldsymbol{\rho},0)$ and $\mathbf{w}_{d,\boldsymbol{\rho}}(\boldsymbol{\rho})$,  
\begin{align}
u_h&=-\frac{1}{2}\left[a_{11}\mathbf{w}^2_{\boldsymbol{\rho}}(\boldsymbol{\rho})+2a_{12}\mathbf{w}_{\boldsymbol{\rho}}(\boldsymbol{\rho})\cdot\mathbf{E}_{\boldsymbol{\rho}}(\boldsymbol{\rho},0)+a_{22}\mathbf{E}^2_{\boldsymbol{\rho}}(\boldsymbol{\rho},0) \right. 
\nonumber \\ &\qquad {} \left. +a_d\mathbf{w}^2_{d,\boldsymbol{\rho}}(\boldsymbol{\rho}) \right].  
\label{endenuhh2}
\end{align}
Inserting this $u_h$ expression into $\ddot{\mathbf{w}}_{\boldsymbol{\rho}}(\boldsymbol{\rho})=-\nabla _{\mathbf{w}}u_h$, $\boldsymbol{\mathcal{P}}(\boldsymbol{\rho})=-\nabla _{\mathbf{E}}u_h$ and $\ddot{\mathbf{w}}_{d,\boldsymbol{\rho}}(\boldsymbol{\rho})=-\nabla _{\mathbf{w}_d}u_h$, where $\nabla _{\mathbf{w}}$ denotes $(\frac{\partial}{\partial w_x},\frac{\partial}{\partial w_y})$, for instance, and the vector subscripts $\mathbf{w}$ and $\mathbf{E}$ and $\mathbf{w}_d$ of the del $\nabla$ are shorthand notations for in-plane vectors $\mathbf{w}_{\boldsymbol{\rho}}(\boldsymbol{\rho})$, $\mathbf{E}_{\boldsymbol{\rho}}(\boldsymbol{\rho},0)$ and $\mathbf{w}_{d,\boldsymbol{\rho}}(\boldsymbol{\rho})$, one  rederives the lattice equations (\ref{eomw1co}), (\ref{bigP1co}) and (\ref{eomwss1co}). Note that,   
similar to the bulk case \cite{Born:1954}, it is not {\it a priori} obvious that an energy density of such a simple form should exist for ML MoS$_2$, in particular when considering $\mathbf{E}$ is the macroscopic field -- not simply an externally applied  field.  

With the relation $c_{12}=c_{21}$, similarly, an areal energy density associated with the out-of-plane optical vibrations can be defined,  
\begin{align}
u_v&=-\frac{1}{2}\left[c_{11}\mathbf{w}^2_z(\boldsymbol{\rho})+2c_{12}\mathbf{w}_z(\boldsymbol{\rho})\cdot\mathbf{E}_z(\boldsymbol{\rho},0)+c_{22}\mathbf{E}^2_z(\boldsymbol{\rho},0) \right. 
\nonumber \\ &\qquad {} \left. +c_d\mathbf{w}^2_{d,z}(\boldsymbol{\rho})\right],   
\label{endenuvv2}
\end{align}
from which the lattice equations (\ref{eomw1zco}), (\ref{bigP1zco}) and (\ref{eomwss1zco}) can be rederived through $\ddot{w}_z=-\partial u_v/\partial w_z$, $\mathcal{P}_z=-\partial u_v/\partial E_z$ and $\ddot{w}_{d,z}=-\partial u_v/\partial w_{d,z}$ [$w_z=w_z(\boldsymbol{\rho})$, $E_z=E_z(\boldsymbol{\rho},0)$ and $w_{d,z}=w_{d,z}(\boldsymbol{\rho})$].

\subsection{In-plane and out-of-plane polar optical modes}

The nonpolar optical modes due to the relative motion of the negative ions are dispersionless [Eqs.~(\ref{add1}) and (\ref{cdd1z})] so we focus on the polar modes, i.e., LO$_2$, TO$_2$ and ZO$_2$ (but for simplicity the index subscripts 2 are omitted hereafter, causing no confusion), corresponding to the motion of the positive ions relative to the negative ions in the 2D polar crystals.  
The in-plane optical modes can be obtained from Eqs.~(\ref{eomw1}) and (\ref{bigP1}) plus the equation of electrostatics 
$\nabla\cdot(\mathbf{E}+4\pi\mathbf{P})=0$, where $\mathbf{P}$ is the polarization $\mathbf{P}=\boldsymbol{\mathcal{P}}\bar{\delta}(z)$, and 
 $\mathbf{E}$ is an electrostatic field $\mathbf{E}=-\nabla\phi$. Let $\mathbf{w}(\boldsymbol{\rho})=\mathbf{w}_0e^{i\mathbf{k}\cdot\boldsymbol{\rho}}$ and electric potential  $\phi(\boldsymbol{\rho},z)=\varphi(z)e^{i\mathbf{k}\cdot\boldsymbol{\rho}}$ (time dependence $e^{-i\omega t}$ is omitted).  Expressing $\mathbf{E}$ in terms of $\varphi$ one has the in-plane component of $\mathbf{E}$ at $z=0$ in the ML, $\mathbf{E}_{\boldsymbol{\rho}}(\boldsymbol{\rho},0)=-i\mathbf{k}\varphi(0)e^{i\mathbf{k}\cdot\boldsymbol{\rho}}$. Applying divergence $\nabla_{\boldsymbol{\rho}}$ to  Eq.~(\ref{bigP1}), one finds  the polarization charge density through $-\nabla\cdot\mathbf{P}=-\bar{\delta}(z)\nabla_{\boldsymbol{\rho}}\cdot\boldsymbol{\mathcal{P}}$ and then    
has Poisson's equation,  
\begin{equation}
\nabla^2\phi(\boldsymbol{\rho},z)=4\pi\bar{\delta}(z)[a_{21}i\mathbf{k}\cdot\mathbf{w}(\boldsymbol{\rho})+a_{22}k^2\varphi(0)e^{i\mathbf{k}\cdot\boldsymbol{\rho}}]. 
\label{poi1}
\end{equation}

Inserting the expansions of $\varphi(z)$ [Eq.~(\ref{vphx})] and $\delta(z-z_{\nu})$ [Eq.~(\ref{delx})] into this equation, one obtains  
\begin{equation}
\hat{\varphi}(q)=-\left[a_{21}i\mathbf{k}\cdot\mathbf{w}_0+a_{22}k^2\varphi(0)\right]\frac{2}{3(k^2+q^2)}\big(1+2\cos\frac{qd}{2}\big). 
\label{phiq}
\end{equation}
From the integral  $\varphi(0)=\int_{-\infty}^{\infty}\hat{\varphi}(q)dq$, one finds $\varphi(0)$,  
\begin{equation}
\varphi(0)=-\frac{2\pi a_{21}\gamma_ki\mathbf{k}\cdot\mathbf{w}_0}{k(1+2\pi a_{22}k\gamma_k)},   
\label{vph0a}
\end{equation}
and then obtains the electric field in the ML,  
\begin{equation}
\mathbf{E}_{\boldsymbol{\rho}}(\boldsymbol{\rho},0)=-\frac{2\pi a_{21}\gamma_k\mathbf{k}}{k(1+2\pi a_{22}k\gamma_k)}\mathbf{w}\cdot\mathbf{k},     
\label{Eroz0}
\end{equation}
where $\gamma_k=(1+2e^{-kd/2})/3$. 

For a normal mode with wavevector $\mathbf{k}$,  Eq.~(\ref{eomw1}) becomes 
\begin{equation}
(-\omega^2-a_{11})\mathbf{w}(\boldsymbol{\rho})=a_{12}\mathbf{E}_{\boldsymbol{\rho}}(\boldsymbol{\rho},0).   
\label{eow1b}
\end{equation}

Expression (\ref{Eroz0}) admits two possibilities for $\mathbf{w}\cdot\mathbf{k}$, (i) $\mathbf{w}\cdot\mathbf{k}=0$, or (ii) $\mathbf{w}\cdot\mathbf{k}\ne 0$. 
In case (i) $\mathbf{w}\perp\mathbf{k}$ so the normal modes are transverse waves. According to Eqs.~(\ref{phiq}) and (\ref{vph0a}), $\hat{\varphi}(q)=0$. Then from Eq.~(\ref{vphx}) one finds $\varphi(z)=0$, and thus the macroscopic field vanishes,  $\mathbf{E}(\mathbf{r})=0$. Therefore the frequency of the TO mode is 
\begin{equation}
\omega_t=\omega_0=\sqrt{-a_{11}}=\sqrt{\frac{2(K_1-e_ae_BQ_1)}{\bar{m}}},       
\label{wto2a}
\end{equation}
independent of wavevector; that is, the long-wavelength TO modes are dispersionless.

In case (ii) the electric field $\mathbf{E}_{\boldsymbol{\rho}}(\mathbf{r})$ is nonzero, and evidently the vectors $\mathbf{w}(\boldsymbol{\rho})$, $\mathbf{E}_{\boldsymbol{\rho}}(\mathbf{r})$, $\mathbf{P}(\mathbf{r})$ associated with  the mode are all longitudinal, i.e., $\mathbf{w}(\boldsymbol{\rho})\parallel\mathbf{E}_{\boldsymbol{\rho}}(\mathbf{r})\parallel\mathbf{P}(\mathbf{r})\parallel\mathbf{k}$. Inserting  expression (\ref{Eroz0}) into Eq.~(\ref{eow1b}) yields the frequency of the longitudinal optical (LO) mode,   
\begin{align}
\omega_l(k)&=\Big(-a_{11}+\frac{2\pi a_{21}^2k\gamma_k}{1+2\pi a_{22}k\gamma_k}\Big)^{1/2}
\nonumber \\ &\qquad {} =\Big[\omega_0^2+\frac{2\pi e_B^2k\gamma_k}{\bar{m}s(1+2\pi\chi_ek\gamma_k)}\Big]^{1/2},      
\label{wlo2a}
\end{align}
where the $a$-coefficients are changed to $e_B$, $\chi_e$, $\omega_0$ by  Eqs.~(\ref{a22alf}), (\ref{a21b}) and (\ref{wto2a}). 
For the long wavelengths, $k\ll 1/a\approx 1/d$, $\gamma_k\approx 1$, one obtains the general LO phonon dispersion relation, 
\begin{equation}
\omega_l(k)=\Big[\omega_0^2+\frac{2\pi e_B^2k}{\bar{m}s(1+2\pi\chi_ek)}\Big]^{1/2},       
\label{wlo2a2b}
\end{equation}
independent of the specific form of $\bar{\delta}(z)$, 
 as the effect of the finite  thickness of ML MoS$_2$ is negligible for the long waves. 
$\omega_l$ increases monotonically with $k$, with an upper bound $\omega_M$ at very large $k$, $\omega_M=\sqrt{\omega_0^2+e_B^2/(\bar{m}s\chi_e)}$. The dispersion relation (\ref{wlo2a2b}) is identical to the analytical expression of Sohier {\it et al}. \cite{Sohier:2017,Zhang:2019b}. 

Expanding $\omega_l(k)$ [Eq.~(\ref{wlo2a2b})] to second order in $k$,  one has 
\begin{equation}
\omega_l(k)=\omega_0+c_lk-\frac{1}{2}c_l(\frac{c_l}{\omega_0}+4\pi \chi_e)k^2,     
\label{wlok0}
\end{equation}
where  $c_l=\pi e_B^2/(\bar{m}s\omega_0)$  \cite{Michel:2009} is the norm of the LO phonon group velocity corresponding to the slope of the dispersion curve at $\Gamma$ \cite{Sohier:2017}. For an arbitrary $k$, the norm of the group velocity,   
\begin{equation}
\omega_l'(k)=c_l\Big[1+\frac{2c_lk/\omega_0}{1+2\pi\chi_ek}\Big]^{-1/2}(1+2\pi\chi_ek)^{-2},     
\label{wlovgka}
\end{equation}
decreases as $k$ becomes larger, showing that the dispersion curve becomes flatter as $k$ increases.
The LO phonon dispersion curves from first-principles calculations  \cite{Sohier:2017} can be described very well using  Eqs.~(\ref{wlo2a2b}) and (\ref{wlovgka}).

The density of states of the LO modes can be obtained from the dispersion Eq.~(\ref{wlo2a2b}),   
\begin{equation}
g_l(\omega)=\frac{c_l\omega_0}{4\pi^4 \chi_e^3}\frac{\omega(\omega^2-\omega_0^2)}{\big(\omega_M^2-\omega^2\big)^3}, \quad \omega_0 \leq \omega < \omega_M.     
\label{doslo1}
\end{equation}

Having $\varphi(0)$ [Eq.~(\ref{vph0a})], one substitutes expression (\ref{phiq}) back into Eq.~(\ref{vphx}) to find the electric potential and then obtains the macroscopic field associated with the LO mode, 
\begin{subequations} 
\begin{equation}
\mathbf{E}_{\boldsymbol{\rho}}(\boldsymbol{\rho},z)=-\frac{2\pi e_B\mathbf{k}}{3\sqrt{\bar{m}s}k(1+2\pi \chi_ek)}\mathbf{w}\cdot\mathbf{k}\sum_{\nu}e^{-k\lvert z-z_{\nu}\rvert}, 
\label{Erolo}
\end{equation}
\begin{equation}
\mathbf{E}_z(\boldsymbol{\rho},z)=-\mathbf{e}_z\frac{2\pi ie_B}{3\sqrt{\bar{m}s}(1+2\pi \chi_ek)}\mathbf{w}\cdot\mathbf{k}\sum_{\nu}\sgn(z-z_{\nu})e^{-k\lvert z-z_{\nu}\rvert},   
\label{Ezlo}
\end{equation}
\end{subequations} 
consistent with  Eqs.~(\ref{Ero1}) and (\ref{Ez1}) above.

This macroscopic field results in a higher LO frequency than the TO frequency at a finite $k$ \cite{Sanchez:2002} with the splitting given in Eq.~(\ref{wlo2a}). No splitting occurs in the limit $k\rightarrow 0$ as the macroscopic field vanishes [Eq.~(\ref{Erolo})], which is different from the case in bulk 2H-MoS$_2$, where LO-TO splitting occurs at the $\Gamma$ point \cite{Cai:2014}.  Thus the transparent expressions (\ref{wto2a}) and (\ref{wlo2a}) describe the degeneracy at $\Gamma$ and the splitting at a finite wavevector of the LO and TO modes, well-known  phenomena of the 2D semiconductors \cite{Sanchez:2002,Wirtz:2003,Michel:2009,Topsakal:2009,Ataca:2011,Cai:2014,Sohier:2017}.

The out-of-plane optical modes can be obtained from Eqs.~(\ref{eomw1z}) and (\ref{bigP1z}) as follows. 
Insert the field in the ML $\mathbf{E}_z(\boldsymbol{\rho},0)=-\mathbf{e}_z\varphi'(0)e^{i\mathbf{k}\cdot\boldsymbol{\rho}}$ into Eq.~(\ref{bigP1z}) and then express the polarization charge density $-\nabla\cdot\mathbf{P}$ in terms of $\mathbf{w}\cdot\mathbf{e}_z$ and $\varphi'(0)$.  The Poisson equation is given by  
\begin{equation}
\nabla^2\phi(\boldsymbol{\rho},z)=4\pi\bar{\delta}'(z)[c_{21}\mathbf{w}(\boldsymbol{\rho})\cdot\mathbf{e}_z-c_{22}\varphi'(0)e^{i\mathbf{k}\cdot\boldsymbol{\rho}}]. 
\label{poiz1}
\end{equation}

When the expansions of $\varphi(z)$ and $\delta(z-z_{\nu})$ [Eqs.~(\ref{vphx}) and (\ref{delx})] are inserted, one finds $\hat{\varphi}(q)$,  
\begin{equation}
\hat{\varphi}(q)=-\left[c_{21}\mathbf{w}_0\cdot\mathbf{e}_z-c_{22}\varphi'(0)\right]\frac{2iq}{3(k^2+q^2)}\big(1+2\cos\frac{qd}{2}\big),  
\label{phiqz}
\end{equation}
and after the integration $\varphi'(z)=\int_{-\infty}^{\infty}iq\hat{\varphi}(q)e^{iqz}dq$, 
one obtains 
\begin{equation}
\varphi'(z)=4\pi\Big[c_{21}\mathbf{w}_0\cdot\mathbf{e}_z-c_{22}\varphi'(0)\Big]\Big[\bar{\delta}(z)-\frac{1}{6}k\sum_{\nu}e^{-k\lvert z-z_{\nu}\rvert}\Big].    
\label{phi1za}
\end{equation}
This expression has a $\bar{\delta}(z)$ term (divergent at $z_{\nu}=0, \pm d/2$), because the atomic layers are treated as geometric planes where the ionic charge distribution and polarization density $\mathbf{P}$ have a $\delta(z-z_{\nu})$ form. Now we approximate $\delta(z-z_{\nu})$ by a Gaussian distribution with a small effective thickness $\varepsilon$ 
($\varepsilon\rightarrow 0$ ), $\delta_{\varepsilon}(z-z_{\nu})=\frac{1}{\sqrt{\pi}\varepsilon}e^{-(z-z_{\nu})^2/\varepsilon^2}$,  
as in previous theoretical study \cite{Michel:2009} and first-principles calculations \cite{Sohier:2017}. 
Taking $z=0$ in Eq.~(\ref{phi1za}) then one finds $\varphi'(0)$, 
\begin{equation}
\varphi'(0)=\frac{c_{21}}{\varepsilon/\zeta_k+c_{22}}\mathbf{w}_0\cdot\mathbf{e}_z,   
\label{phi1z0b}
\end{equation}
where 
\begin{equation}
\zeta_k=\frac{4\sqrt{\pi}}{3}\Big[1+2e^{-(d/2\varepsilon)^2}-\sqrt{\pi}\frac{k\varepsilon}{2}\big(1+2e^{-kd/2}\big)\Big].    
\label{zetak}
\end{equation}

The electric field in the ML $\mathbf{E}_z(\boldsymbol{\rho},0)$ follows,   
\begin{equation}
\mathbf{E}_z(\boldsymbol{\rho},0)=-\frac{c_{21}}{\varepsilon/\zeta_k+c_{22}}\mathbf{w}=-\frac{e_B^\prime}{\sqrt{\bar{m}s}(\varepsilon/\zeta_k+\chi_e^\prime)}\mathbf{w}. 
\label{Ezro0b1}
\end{equation}
Substituting this field into the equation of motion (\ref{eomw1z}), one obtains the frequency of the out-of-plane mode,
\begin{equation}
\omega_z(k)=\Big[\omega_0'^2+\frac{e_B^{\prime 2}}{\bar{m}s(\varepsilon/\zeta_k+\chi_e^\prime)}\Big]^{1/2}.     
\label{wzo2a}
\end{equation}

When the broadening $\varepsilon$ is similar to the dimensions of a lattice cell,  expression (\ref{zetak}) is simplified to a constant $\zeta=4\sqrt{\pi}[1+2e^{-(d/2\varepsilon)^2}]/3$,  
and the phonon frequency becomes independent of wavevector;   
for ML MoS$_2$ when $\varepsilon \ll 4\sqrt{\pi}\chi_e^\prime/3$ the frequency reduces to 
\begin{equation}
\omega_z=\Big(\omega_0'^2+\frac{e_B^{\prime 2}}{\bar{m}s\chi_e^\prime}\Big)^{1/2}.     
\label{wzo2b3}
\end{equation}
The EP of ions (i.e., $\chi_e^\prime\ne 0$ ) ensures a finite frequency, otherwise $\omega_z$ [Eq.~(\ref{wzo2a})] becomes very large and even 
$\omega_z\rightarrow \infty$ [Eq.~(\ref{wzo2b3})] when $\chi_e^\prime$ is neglected in the rigid ion model (i.e., without EP).

With $\varphi'(0)$ [Eq.~(\ref{phi1z0b})] then one can get $\varphi'(z)$ from  Eq.~(\ref{phi1za}) and also obtain $\varphi(z)$ after substituting the $\hat{\varphi}(q)$ expression (\ref{phiqz}) into Eq.~(\ref{vphx}). It follows that the macroscopic field associated with the out-of-plane polar mode is given by the following expressions [consistent with Eqs.~(\ref{Ez2}) and (\ref{Ero2})], 
\begin{subequations} 
\begin{equation}
\mathbf{E}_{\boldsymbol{\rho}}(\boldsymbol{\rho},z)=\frac{-2i\pi e_B^\prime\mathbf{k}}{3\sqrt{\bar{m}s}(1+\chi_e^\prime 4\sqrt{\pi}/3\varepsilon)}\mathbf{w}\cdot\mathbf{e}_z\sum_{\nu}\sgn(z-z_{\nu})e^{-k\lvert z-z_{\nu}\rvert},  
\label{Erozo9}
\end{equation}
\begin{equation}
\mathbf{E}_z(\boldsymbol{\rho},z)=\frac{-4\pi e_B^\prime}{\sqrt{\bar{m}s}(1+\chi_e^\prime 4\sqrt{\pi}/3\varepsilon)}\mathbf{w}\Big[\bar{\delta}(z)-\frac{1}{6}k\sum_{\nu}e^{-k\lvert z-z_{\nu}\rvert}\Big].    
\label{Ezzo9}
\end{equation}
\end{subequations} 
Note that the in-plane component, antisymmetric with respect to the Mo layer, is negligible at very small $k$ and $\varepsilon$.

\subsection{2D lattice dielectric susceptibility and dielectric function}

The in-plane lattice susceptibility can be derived from Eqs.~(\ref{eomw1}) and (\ref{bigP1}) by considering  periodic solutions $\mathbf{E}$, $\boldsymbol{\mathcal{P}}$, $\mathbf{W}$ $\propto e^{-i\omega t}$ due to an electric disturbance with frequency $\omega$.  The susceptibility, $\chi=\boldsymbol{\mathcal{P}}(\boldsymbol{\rho})/\mathbf{E}_{\boldsymbol{\rho}}(\boldsymbol{\rho},0)$, is given by 
\begin{equation}
\chi(\omega)=a_{22}-\frac{a_{12}a_{21}}{\omega^2+a_{11}}=\chi_e+\frac{e_B^2}{\bar{m}s(\omega_0^2-\omega^2)},          
\label{chiwp1}
\end{equation}
where the $\omega$-terms are due to lattice vibration.  From Eq.~(\ref{chiwp1}) the static susceptibility is  
\begin{equation}
\chi_0=\chi_e+\frac{e_B^2}{\bar{m}s\omega_0^2}.        
\label{chiwps0}
\end{equation}
Using Eq.~(\ref{chiwps0}), one can express $a_{12}$ [Eq.~(\ref{a12})] in terms of the 2D susceptibilities as 
\begin{equation}
a_{12}=a_{21}=\omega_0\sqrt{\chi_0-\chi_e},         
\label{a12chi09}
\end{equation}
which has the same form as its 3D counterpart $b_{12}$ ($b_{21}$) that is expressed in terms of the 3D susceptibilities or dielectric constants \cite{Born:1954}. 

Further the 2D dynamical susceptibility $\chi(\omega)$ can be conveniently expressed in terms of the oscillator frequency $\omega_0$ and the high-frequency and static susceptibilities $\chi_e$, $\chi_0$ as
\begin{equation}
\chi(\omega)=\chi_e+\frac{\chi_0-\chi_e}{1-\omega^2/\omega_0^2}~,       
\label{chiwp2}
\end{equation}
which is similar to its counterpart of bulk crystals \cite{Born:1954}.   

The 2D susceptibility for vertical polarization, $\chi'=\boldsymbol{\mathcal{P}}(\boldsymbol{\rho})/\mathbf{E}_z(\boldsymbol{\rho},0)$, is derived from Eqs.~(\ref{eomw1z}) and (\ref{bigP1z}), 
\begin{equation}
\chi'(\omega)=c_{22}-\frac{c_{12}c_{21}}{\omega^2+c_{11}}=\chi_e'+\frac{e_B'^2}{\bar{m}s(\omega_0'^2-\omega^2)},          
\label{chiwz1}
\end{equation}
and the static susceptibility is       
\begin{equation}
\chi_0'=\chi_e'+\frac{e_B'^2}{\bar{m}s\omega_0'^2}.        
\label{chiwzs0}
\end{equation}
Using this, one can rewrite $c_{12}$ [Eq.~(\ref{c12z})] in terms of the 2D susceptibilities as 
\begin{equation}
c_{12}=c_{21}=\omega_0'\sqrt{\chi_0'-\chi_e'},         
\label{c12chi09}
\end{equation}
and also transform the dynamical susceptibility $\chi'(\omega)$ as   
\begin{equation}
\chi'(\omega)=\chi_e'+\frac{\chi_0'-\chi_e'}{1-\omega^2/\omega_0'^2}.      
\label{chiwz2}
\end{equation}

On eliminating $e_B'^2/(\bar{m}s)$ in Eq.~(\ref{wzo2b3}) with expression (\ref{chiwzs0})  one obtains a simple relation 
\begin{equation}
\frac{\omega_z^2}{\omega_0'^2}=\frac{\chi_0'}{\chi_e'}.      
\label{wz02chz0}
\end{equation}

The lattice DF can be derived by introducing a test charge to calculate the total potential in the electrostatic approximation as in Ref.\cite{Cudazzo:2011}. To find the potential let the test charge density function $\sigma$ have the form $\sigma=\sigma_0e^{i\mathbf{k}\cdot\boldsymbol{\rho}}\delta(z-z_a)$, where $z_a$ is the $z$-coordinate of the test charge not necessarily confined in the ML, and time dependence $e^{-i\omega t}$ is omitted for clearness. Now the equation of electrostatics is $\nabla\cdot(\mathbf{E}+4\pi\mathbf{P})=4\pi\sigma$, $\mathbf{E}$ being the total field of the test charge and polarization charge. In general the field of the test charge is nonzero in the ML,  and thus the lattice responds generating in-plane [Eqs.~(\ref{eomw1co}) and (\ref{bigP1co})] and out-of-plane [Eqs.~(\ref{eomw1zco}) and (\ref{bigP1zco})] vibrations, with all the quantities varying as $e^{i(\mathbf{k}\cdot\boldsymbol{\rho}-\omega t)}$.  As the induced potential associated with the out-of-plane polarization is zero at the centre of the ML and antisymmetric with respect to the Mo layer [refer to Eq.~(\ref{Pophez}) above], the out-of-plane motion makes no contribution to the DF \cite{Zhang:2019b} and we only need to consider the in-plane motion,  taking the dielectric polarization $\mathbf{P}=\boldsymbol{\mathcal{P}}_{\boldsymbol{\rho}}\bar{\delta}(z)$, with $\boldsymbol{\mathcal{P}}_{\boldsymbol{\rho}}=\chi(\omega)\mathbf{E}_{\boldsymbol{\rho}}(\boldsymbol{\rho},0)$.

Writing $\mathbf{E}=-\nabla\phi$ with the total potential $\phi(\mathbf{r})=\varphi(z)e^{i\mathbf{k}\cdot\boldsymbol{\rho}}$, one has Poisson's equation, 
\begin{equation}
\nabla^2\phi(\boldsymbol{\rho},z)=-4\pi\left[\sigma_0 e^{i\mathbf{k}\cdot\boldsymbol{\rho}}\delta(z-z_a)+\chi(\omega)\bar{\delta}(z)
\nabla_{\boldsymbol{\rho}}^2\phi(\boldsymbol{\rho},0)\right], 
\label{poions1}
\end{equation}
where $\phi(\boldsymbol{\rho},0)$ is the total potential in the ML \cite{Cudazzo:2011}, $\phi(\boldsymbol{\rho},0)=\varphi(0)e^{i\mathbf{k}\cdot\boldsymbol{\rho}}$.  
Using expansions (\ref{vphx}) and (\ref{delx}) then one finds $\hat{\varphi}(q)$, 
\begin{equation}
\hat{\varphi}(q)=\frac{2}{k^2+q^2}\left[ \sigma_0e^{-iqz_a}-\frac{1}{3}\chi(\omega)k^2\varphi(0)\Big(1+2\cos\frac{qd}{2}\Big) \right].        
\label{vaphqdcp}
\end{equation}

The integration $\varphi(0)=\int_{-\infty}^{\infty}\hat{\varphi}(q)dq$ gives $\varphi(0)$,  
\begin{equation}
\varphi(0)=\frac{2\pi\sigma_0e^{-k\lvert z_a\rvert}/k}{1+2\pi k\gamma_k\chi(\omega)}~.       
\label{vaphz0dc}
\end{equation}

Evidently the numerator multiplied by $e^{i\mathbf{k}\cdot\boldsymbol{\rho}}$ is the electric potential in the ML of the test charge, and therefore the DF of the 2D lattice is the denominator,  
\begin{equation}
\epsilon(k,\omega)=1+2\pi k\gamma_k\chi(\omega)=1+2\pi k\gamma_k\left[\chi_e+\frac{e_B^2}{\bar{m}s(\omega_0^2-\omega^2)}\right].      
\label{df2D}
\end{equation}

The DF $\epsilon(k,\omega)$ is a longitudinal DF due only to the LO vibrations (i.e., no contribution from the TO or ZO vibrations), because 
 the polarization charge density 
for the in-plane motion is given by $-\bar{\delta}(z)\nabla_{\boldsymbol{\rho}}\cdot\boldsymbol{\mathcal{P}}_{\boldsymbol{\rho}}(\boldsymbol{\rho})=-\bar{\delta}(z)[a_{21}-a_{22}(\omega^2+a_{11})/a_{12}]i\mathbf{k}\cdot\mathbf{w}_{\boldsymbol{\rho}}\ne 0$, and also $\mathbf{w}_{\boldsymbol{\rho}}\parallel\boldsymbol{\mathcal{P}}_{\boldsymbol{\rho}}\parallel\mathbf{E}_{\boldsymbol{\rho}}(\boldsymbol{\rho},0)\parallel\mathbf{k}$. Without test charge ($\sigma=0$), a finite electric field $\mathbf{E}_{\boldsymbol{\rho}}(\boldsymbol{\rho},0)$ and potential $\phi(\boldsymbol{\rho},0)$ occur due to the LO vibrations, and from Eq.~(\ref{vaphz0dc}) therefore the LO modes are the solutions to 
\begin{equation}
\epsilon(k,\omega)=0,       
\label{dfwl0}
\end{equation}
which is very similar to the bulk crystal result \cite{Born:1954}.  
Combining expressions (\ref{df2D}) and (\ref{dfwl0}) yields LO phonon dispersion $\omega_l(k)$ indeed identical with expression (\ref{wlo2a}).

The static and high-frequency DFs are $\epsilon_0(k)=1+2\pi\chi_0k\gamma_k$, and $\epsilon_{\infty}(k)=1+2\pi\chi_ek\gamma_k$, respectively. 
The lattice DF of the 2D crystal can be transformed as 
\begin{equation}
\epsilon(k,\omega)=\epsilon_{\infty}(k)\frac{\omega^2-\omega_l^2(k)}{\omega^2-\omega_t^2},       
\label{df2D2}
\end{equation}
i.e., in a very similar form to the lattice DF of bulk crystals \cite{Born:1954}, the difference being that here the LO phonon frequency and DFs are both wavevector-dependent. 
Expression (\ref{df2D2}) shows that the TO phonon frequency $\omega_t$ is a pole of $\epsilon(k,\omega)$ whereas at LO phonon frequencies $\omega_l(k)$ the DF is zero. Further in bulk there is the Lyddane--Sachs--Teller (LST) relation  \cite{Lyddane:1941}, $\omega_l^2/\omega_t^2=\epsilon_0/\epsilon_{\infty}$;  
for the 2D crystal a similar relation can be obtained from expression (\ref{df2D2}), 
\begin{equation}
\frac{\omega_l^2(k)}{\omega_t^2}=\frac{\epsilon_0(k)}{\epsilon_{\infty}(k)}.        
\label{LST2}
\end{equation}

The extended LST relation (\ref{LST2}) shows a clear physical significance: it connects the phonon frequencies with the static and high-frequency DFs; given the former, then ratio of the latter is known, and vice versa. Dispersion occurs in the 2D crystal with the polar LO mode and DFs strongly dependent on the wavevector. Knowing the LO phonon dispersion $\omega_l(k)$ [$\omega_l(0)=\omega_t$], for instance, from an experimental measurement or theoretical calculation, then the ratio between the two DFs $\epsilon_0(k)$ and $\epsilon_{\infty}(k)$ is obtained.  Meanwhile, as the LO and TO mode frequencies differ due solely to the macroscopic field, the relation is very useful for evaluating the phonon frequency increase  brought by the field. 
Similarly, for the out-of-plane motion the frequency--susceptibility relation (\ref{wz02chz0}) measures the effect of the macroscopic field on the phonon frequency.

For the long wavelengths, taking $\gamma_k=1$, Eq.~(\ref{df2D}) becomes a general DF expression
independent of the specific form of $\bar{\delta}(z)$; it is also the same as the DF $\epsilon(k,\omega)$ of ML hBN \cite{Zhang:2019b}, and accordingly the $\epsilon_0(k)$ and $\epsilon_{\infty}(k)$ expressions are simplified to the DF formula for a normal 2D dielectric deduced by Cudazzo {\it et al}. \cite{Cudazzo:2011}.

\section{Numerical results and further discussions}

\subsection{Various lattice dynamical properties of ML TMDs}

In what follows we confine ourselves to in-plane motion, and specifically the polar LO and TO modes, corresponding to the vibrations of the positive ions relative to the negative ions. These modes at the zone center are denoted as $E'$; for ML MoS$_2$, for instance, the first-principles calculations by density-functional perturbation theory (DFPT) find  
$\omega(E')$=380.2 cm$^{-1}$ in Ref.\cite{Ataca:2011} and $\omega(E')$=402.7 cm$^{-1}$ in Ref.\cite{Cai:2014}. 
In a recent study \cite{Zhang:2019b} we found that the atomic polarizability of the  unit cell of ML hBN is significantly reduced compared to the total free-atom polarizablity of B and N. The polarizabilities $\alpha_1$ and $\alpha_2$ of the constituent atoms in ML TMDs may differ significantly from the free-atom values and are unknown quantities so the values of the $a$-coefficients can not be determined with expressions such as Eqs.~(\ref{a21}) and (\ref{a22}). 
A set of three values such as $\omega_0$, $e_B$, $\chi_e$ can be obtained from first-principles calculations and therefore the three mutually independent $a$-coefficients $a_{11}$, $a_{12}$ (or $a_{21}$), $a_{22}$ of the lattice equations can be determined through the expressions (\ref{a22alf}), (\ref{a11}) and (\ref{a12}). 
Furthermore, as the three expressions (\ref{a22eB}), (\ref{wto2a}) and (\ref{chiwps0}) relate the three quantities of a ML TMD, namely the two {\it macroscopic} susceptibilities $\chi_e$, $\chi_0$ and the collective vibration frequency $\omega_0$ to the three {\it microscopic} quantities, i.e., the static effective charge $e_a$, the Born charge $e_B$ and the spring force constant $K_1$, 
one can calculate $e_a$, $K_1$, $\chi_0$ using these expressions.  
Of the four microscopic quantities $\alpha_1$, $\alpha_2$, $e_a$, $K_1$ of our dipole lattice model on which the $a$-coefficients originally depend [refer to  Eqs.~(\ref{a21}), (\ref{a22}), (\ref{eB}), (\ref{Ke}) and (\ref{a11})], only the two atomic  polarizabilities $\alpha_1$, $\alpha_2$ are unknown. The adoption of the set of known quantities $\omega_0$, $e_B$, $\chi_e$ calculated from first principles facilitates the use of the deduced equations by circumventing the unknowns $\alpha_1$ and $\alpha_2$, as we shall see below.

The atomic polarizabilities $\alpha_1$ and $\alpha_2$ can not be determined from Eqs.~(\ref{a22}) and (\ref{eB}) even when $e_B$ and $\chi_e$ are known, 
because $e_B$ and $\chi_e$ are related via Eq.~(\ref{a22eB}) and not independent of each other. Interestingly, 
 the sum $\alpha=\alpha_1+2\alpha_2$, i.e., the total atomic polarizability of the unit cell is limited in an interval (given $\chi_e$), which can be determined using the Clausius-Mossotti relation (\ref{a22}).  Changing variable $\alpha_2$ to $\alpha_1$ for a $\alpha$ transforms Eq.~(\ref{a22}) into a quadratic equation in variable $\alpha_1$. Then follow the lines of the derivation given in Ref.\cite{Zhang:2019b} (Appendix A). By requiring the discriminant $\Delta_{\alpha}\geq 0$ as well as $\alpha_1\geq 0$ and $\alpha_2\geq 0$, one obtains the interval for $\alpha$, 
\begin{align}
&\frac{1}{Q_0+1/\chi_es}\leq\alpha 
\nonumber \\ & \leq\frac{3Q_0+Q_2-4Q_1}{Q_0(Q_0+Q_2)-2Q_1^2+(3Q_0+Q_2-4Q_1)/\chi_es}. 
\label{alfintv}
\end{align}

In a recent study \cite{Sohier:2017} the LO phonon dispersion of a ML TMD or hBN is given by $\omega_l^2=\omega_0^2+\mathcal{S}k/(1+r_{eff}k)$, where the $\mathcal{S}$ parameter relates to the Born charge via $\mathcal{S}=2\pi e_B^2/(\bar{m}s)$ \cite{Sohier:2017,Zhang:2019b}, and $r_{eff}$ is an effective screening length, given by by $r_{eff}=\epsilon_pt/2$ with an effective medium model \cite{Sohier:2016,Sohier:2017}, $\epsilon_p$ and $t$ being effective dielectric constant and effective thickness of the ML material. Both parameters $\mathcal{S}$ and $r_{eff}$ are computed by first-principles calculation \cite{Sohier:2017}. 
A comparison with Eq.~(\ref{wlo2a2b}) gives the screening length $r_{eff}=2\pi\chi_e$.  
Using the $\mathcal{S}$ and $r_{eff}$ values (listed in Table 1 of Ref.\cite{Sohier:2017} together with $\omega_0$) for a ML TMD or hBN, one finds $\lvert e_B\rvert$ and $\chi_e$, and adding $\omega_0$, one has the values for the set of three quantities ($\lvert e_B\rvert$, $\chi_e$, $\omega_0$), and further calculates the quantities $\chi_0$, $\lvert e_a\rvert$, $K_1$ together with the interval of $\alpha$, as detailed above. 
The result is presented in Table~\ref{table:1} for five ML TMDs and ML hBN for comparison.

In Table~\ref{table:1} we give the absolute values of the Born charges as 
their signs cannot be determined from the $\mathcal{S}$ values of Ref.\cite{Sohier:2017}.  Note that 
the effective charges $e_B$ in ML TMDs have been calculated in first-principles studies, but both positive ($e_B>0$) \cite{Ataca:2011} and negative 
($e_B<0$) \cite{Michel:2017,Danovich:2017} values have been found. 
Considering that the key quantities such as $\omega_l(k)$ [Eq.~(\ref{wlo2a2b})] and $\chi_0$ [Eq.~(\ref{chiwps0})] depend on $e_B^2$, and $e_B$ pairs with $e_a$ in expressions (\ref{a22eB}) and (\ref{wto2a}) (recall the pair have the same sign, i.e., $e_Be_a=\lvert e_Be_a\rvert$), with the absolute values we still find some interesting results.  First, 
Comparing the $\chi_e$ and $\chi_0$ values of the ML TMDs, we see that the susceptibility is only slightly increased when the lattice vibrational contribution is included: there is a  0.3\% increase for ML WS$_2$, 9\% increase for ML MoTe$_2$ and 1-3\% increase for the other ML TMDs, and these increases are smaller than the 37\% susceptibility increase of ML hBN.   
The 2D susceptibility is a key parameter of the effective 2D interaction in ML TMDs, i.e., the Keldysh potential  \cite{Keldysh:1978,Cudazzo:2011}, for which either $\chi_e$ or $\chi_0$ can be used, and as the above result indicates, they will not cause much difference to the evaluation of such 2D interaction.  Second, the Born charge values are quite small in the ML TMDs, and they are all smaller than 2$e$ and the $e_B$ value of ML hBN except for ML MoTe$_2$ which has a Born charge of 3.1$e$, thus resulting in a small vibrational contribution to the susceptibility. All these $\lvert e_B\rvert$ values are very close to recent density functional theory (DFT)  \cite{Danovich:2017} and DFPT \cite{Michel:2017} calculations (compare to Table I of Ref.\cite{Danovich:2017} and Table II of Ref.\cite{Michel:2017}) and are within a 7\% of deviation except for ML WS$_2$ with a 9\% deviation in Born charge. 
Compared to 2D hBN, in ML TMDs the ions also carry a small static effective charge. For ML MoS$_2$, for instance, the charge on the Mo ions is 0.22$e$, less than half the charge on the B ions of ML hBN. No $e_a$ values have been reported for ML TMDs so far but in bulk 2H-MoS$_2$ first-principles calculation in conjunction with Mulliken analysis yields a charge transfer of 0.43 electrons from one Mo atom to two S atoms \cite{Ataca:2011}. Third, in a ML TMD the atomic polarizability of the unit cell $\alpha$ has a limited range of values calculated with inequality (\ref{alfintv}), and similar to ML hBN these values are much smaller than the values of the total free-atom polarizability $\alpha_f$ (ninth column), $\alpha_f=\alpha_{1,f}+2\alpha_{2,f}$ for ML TMDs and $\alpha_f=\alpha_{1,f}+\alpha_{2,f}$ for ML hBN, where the free-atom polarizabilities of the constituent atoms $\alpha_{1,f}$ and $\alpha_{2,f}$ are taken from Ref. \cite{PScollect:2006}. For ML MoS$_2$, for instance, the upper and lower bounds of $\alpha$ are 43\% and 72\% smaller than the free-atom polarizability $\alpha_f$, respectively. The numerical result also shows  that for each ML TMD the lower bound value of $\alpha$ is about half of its upper bound value. In addition, $\alpha$'s two bounds, $\alpha_f$ and $\chi_e$ vary with the ML material in quite a similar manner--the atoms that have a larger atomic polarizability $\alpha_f$ form a ML TMD with a larger dielectric susceptibility $\chi_e$. Fourth, a smaller spring force constant $K_1$ combined with a larger reduced mass $\bar{m}$ makes the ML TMDs have a lower mode frequency $\omega_0$ 
than 2D hBN. Five, as the group velocity $c_l$ corresponds to the slope of the LO phonon dispersion curve at $\Gamma$, a small $c_l$ for ML WS$_2$ means its flat phonon dispersion while a large $c_l$ for ML MoTe$_2$ corresponds to a steep slope of the dispersion near $\Gamma$.

\subsection{Local field and polarizable ion effects in ML TMDs}

Knowing the values of the microscopic quantities now we use them to evaluate the EP of ions and LFEs on the polar vibrations. 
Neglecting EP of ions ($\alpha_1=\alpha_2=0$), there is no electronic susceptibility, i.e., $\chi_e=0$, and the static susceptibility is also reduced substantially with only a small contribution due to lattice vibration. When LFEs are not accounted for (equivalent to $Q_0=Q_1=Q_2=0$), the high-frequency dielectric susceptibility of ML TMDs becomes $(\alpha_1+2\alpha_2)/s=\alpha/s$ [refer to Eqs.~(\ref{bigD}), (\ref{a22}) and (\ref{a22alf})], the intervals of which are shown in Table~\ref{table:1} (last column). Clearly these are small values compared to the $\chi_e$ including the LFEs (second column), with the upper bounds being only 13-18\% of $\chi_e$ and the lower bounds only 6-9\% of $\chi_e$. Without LFEs,  the Born charge becomes smaller, equal to the static effective charge, i.e., $e_B=2e_a$  [refer to Eqs.~(\ref{bigD}) and  (\ref{eB})], which is the same as in the case when ionic EP is neglected because $Q_i$ ($i=0,1,2$) and $\alpha_j$ ($j=1,2$)  form products in the $e_B$'s expression (\ref{eB}). The LO phonon group velocity $c_l$ 
is also decreased substantially, which is only 3-6\% of the $c_l$ value calculated with the polarizable ion model (PIM) including the LFEs. 
The intrinsic oscillator frequency $\omega_0$ is determined by $K_1-K_e$ [Eq.~(\ref{a11})] rather than by $K_1$ alone, i.e., $\omega_0^2=2(K_1-e_ae_BQ_1)/\bar{m}$, and thus excluding LFEs raises the $\omega_0$ value with a percentage increase given by $1/(K_1/K_e-1)$. We find that the percentage increase in $\omega_0^2$ of ML TMDs is between 0.2\% for ML WS$_2$ and 7.5\% for ML MoTe$_2$--much smaller than the 31\% increase for ML hBN--indicating that the LFEs have only a small influence on the  intrinsic frequency $\omega_0$ of ML TMDs.

Further the polarizable ion and local field effects on LO phonon dispersion can be evaluated using the values of Table~\ref{table:1}. In the PIM including the LFEs, the LO phonon frequency $\omega_l(k)$ can be calculated from Eq.~(\ref{wlo2a2b}) using $e_B$, $\chi_e$ in Table~\ref{table:1} and $\omega_{TO}$ in Table 1 of Ref.\cite{Sohier:2017}, or equally by the dispersion formula of Sohier {\it et al.} \cite{Sohier:2017} using $\mathcal{S}$, $r_{eff}$, $\omega_{TO}$ in their Table 1.  
When LFEs are neglected in the PIM, $\omega_l(k)$ becomes dependent on the unit-cell atomic polarizability $\alpha$, given by $\omega_l^2(k)=2K_1/\bar{m}+8\pi e_a^2k/[\bar{m}(s+2\pi\alpha k)]$, which was calculated using two values of $\alpha$, i.e., its lower bound $\alpha_l$ and upper bound $\alpha_u$ in Table~\ref{table:1}.
In the rigid ion model (RIM) accounting for the LFEs, the LO phonon dispersion is given by 
$\omega_l^2(k)=2(K_1-2e_a^2Q_1)/\bar{m}+8\pi e_a^2k/(\bar{m}s)$, and when neglecting LFEs the LO phonon dispersion is simplified to $\omega_l^2(k)=2K_1/\bar{m}+8\pi e_a^2k/(\bar{m}s)$. A comparison of these phonon dispersion relations is demonstrated with ML MoS$_2$ and ML MoTe$_2$ in Figs.~\ref{fig2}(a) and \ref{fig2}(b), respectively. We see that the phonon dispersion curves of the two ML TMDs are similar.  The RIM without LFEs yields the highest phonon frequency, which increases linearly with the wavevector in the long wavelength region (top line). The LFEs cause a reduction to the frequencies but the dispersion remains linear (lower dashed line). When the EP of ions is included, the dispersion becomes nonlinear with a smaller slope at a larger wavevector in both cases of excluding (dotted and dot-dashed lines) and including the LFEs (solid line). Although the values of $\alpha_1$ and $\alpha_2$ are unknown, using the intervals of the total $\alpha$ the PIM without LFEs yields phonon frequencies that are limited in a very {\it narrow} range (i.e., dotted and dot-dashed lines), 0.5 cm$^{-1}$ for ML MoS$_2$ and 2 cm$^{-1}$ for ML MoTe$_2$.  
Clearly  
the phonon frequencies are reduced significantly due to the ionic EP and LFEs (solid line), and in particular,  
 there is a steep slope on the small wavevector side corresponding to a substantially increased phonon group velocity. The phonon dispersion curves of both monolayers become flatter at larger wavevectors [refer to Eq.~(\ref{wlovgka})]. A wider frequency range of ML MoTe$_2$ as shown in Fig.~\ref{fig2}, however means that it has a larger LO phonon group velocity $c_l$ than ML MoS$_2$, consistent with the result in Table~\ref{table:1}.

For out-of-plane optical vibrations, the polar mode ($A_2''$ mode) and nonpolar mode ($A_1$ mode) frequencies have been calculated for ML MoS$_2$ using DFPT,  
$\omega(A_2'')$=490.5 cm$^{-1}$ and $\omega(A_1)$=423.9 cm$^{-1}$ in Ref.\cite{Cai:2014}; $\omega(A_2'')$=465.0 cm$^{-1}$ and $\omega(A_1)$=406.1 cm$^{-1}$ in Ref.\cite{Ataca:2011}. So far there has been no report on Born charge or 2D susceptibility, and when these basic quantities are available, the dynamical  properties can be calculated using our expressions in a similar way to the in-plane vibration calculation.



\section{Summary and Conclusions}
We have deduced two sets of three equations [Eqs.~(\ref{eomw1co}), (\ref{bigP1co}), 
(\ref{eomwss1co}) and Eqs.~(\ref{eomw1zco}), (\ref{bigP1zco}), (\ref{eomwss1zco})] to describe the long wavelength in-plane and out-of-plane lattice vibrations, respectively, in monolayer TMDs using a microscopic dipole lattice model accounting for the LFEs and EP self-consistently. 
The two pairs of equations [Eqs.~(\ref{eomw1co}), (\ref{bigP1co}) and  Eqs.~(\ref{eomw1zco}), (\ref{bigP1zco})], which have the same forms as those for ML hBN \cite{Zhang:2019b} and are also similar to Huang's equations for bulk crystals, 
describe the polar optical lattice vibrations,  whereas the other two equations [Eqs.~(\ref{eomwss1co}) and (\ref{eomwss1zco})], in which the effective forces are proportional to the relative ionic displacements and have a similar form to the Hooke's law formula, 
describe the nonpolar optical vibrations in the 2D polar crystals.  
The averaging of the microscopic distributions in the $z$ direction of the dipole moment density makes it possible to define a 2D macroscopic polarization $\boldsymbol{\mathcal{P}}(\boldsymbol{\rho})$ and express the volume polarization of the ML TMDs as a product of $\boldsymbol{\mathcal{P}}(\boldsymbol{\rho})$ and the averaged microscopic distribution $\bar{\delta}(z)$. 
These together with the 2D Lorentz relations are fundamental to deducing the lattice equations from the atomic theory. As the linear coefficients of the equations are related to the quantities that can be obtained from first principles calculation, the lattice equations are very useful for studying the lattice dynamical properties  analytically.  We have also obtained the expressions for the areal energy densities associated with the in-plane and out-of-plane optical vibrations, respectively.  

The lattice equations [Eqs.~(\ref{eomw1co}), (\ref{bigP1co}) or  Eqs.~(\ref{eomw1zco}), (\ref{bigP1zco})] are solved simultaneously with the   
 equation of electrostatics to deduce the polar optical modes. 
Explicit expressions have been obtained for the frequencies, the macroscopic fields associated with the polar LO and ZO modes, and also the LO phonon group velocity and density of states.  
The LO phonon dispersion relation, applicable to both ML TMDS and ML hBN, 
is identical to the analytical expression of Sohier {\it et al}., and it evidently shows that the LO and TO modes are degenerate at $\Gamma$ and split up at finite wavevectors due to the long-range macroscopic field, characteristic of these 2D polar crystals.  
The transparent LO phonon dispersion relation also describes the first-principles calculation results very well: the LO mode frequency increases with wavevector but the dispersion becomes flatter at larger wavevectors.  
The phonon frequency expressions show that apart from the LO branch all the other five optical branches are dispersionless at the long wavelengths. 
It is also found that the ZO phonon frequency is finite due to ionic EP, but  otherwise with the rigid ion model the ZO frequency becomes extremely large.

The in-plane and out-of-plane lattice dielectric susceptibilities have been deduced from the lattice equations. The 2D longitudinal lattice dielectric function $\epsilon(k,\omega)$ has also been derived, allowing the LO phonon dispersion to be rederived from $\epsilon(k,\omega)=0$, similar to the case of bulk crystals. Further, the 2D Lyddane--Sachs--Teller relation (\ref{LST2}) and frequency--susceptibility relation (\ref{wz02chz0}) have been derived for in-plane and out-of-plane vibrations, respectively, which are very useful for evaluating the effects of the macroscopic field on the phonon frequencies.  
 
We have demonstrated an application of the analytical expressions by using them to study the lattice dynamical properties for in-plane vibrations, when knowing a set of three quantities $\omega_0$, $\chi_e$, $e_B$ from first-principles calculation. In ML TMDs except ML MoTe$_2$ and MoSe$_2$ the effective charges are quite small, causing a small vibrational contribution (less than 1.4\%) to the dielectric susceptibility. Although the individual atomic polarizabilities in ML TMDs are unknown,     
the total atomic polarizability of the unit cell is limited in an interval which is obtained from the Clausius-Mossotti relation, and the intervals for ML TMDs and hBN have been calculated and also compared with the total free-atom polarizabilities.  
The LFEs and EP should be included simultaneously; otherwise, neglecting either or both underestimates substantially the calculated properties: the Born charge decreases simply to the value of the static effective charge (which becomes 78\% smaller for ML MoS$_2$, for instance), the dielectric susceptibilities $\chi_e$ and $\chi_0$ are underestimated by over 80\%, and the LO phonon group velocity is  underestimated by over 90\%. With no EP or LFE, the LO modes have  very small dispersion, showing linear dispersion for the two RIMs and a very narrow frequency range (0.5-2.0 eV) for the PIM without LFE, which is distinct from the LO phonon dispersion calculated after including both ionic EP and LFEs.


\begin{acknowledgments}
We acknowledge support from the New Energy and Materials Collaboration project of the School of Physics, and the Natural Science Research Funds (Nos. 419080500175 \& 419080500260) of Jilin University. 
\end{acknowledgments}


\newpage


%

\newpage

\begin{table*}[htbp]
	\begin{center}
		\leavevmode
		\setlength{\tabcolsep}{5pt}
		\renewcommand\arraystretch{1.5}
		\caption{\label{table:1}  Various calculated properties associated with in-plane optical vibrations in five ML TMDs and ML hBN, namely, the high-frequency susceptibility $\chi_e$, the static susceptibility $\chi_0$, the absolute values of the Born charge $e_B$ and the static effective charge $e_1$ of metal ions ($e_1=2e_a$ for ML TMDs, and $e_1=e_a$ for ML hBN), the effective spring force constant $K_1$, the LO phonon group velocity at the $\Gamma$ point $c_l$, the atomic polarizability of the unit cell $\alpha$, the total free-atom polarizability of the constituent atoms $\alpha_f$ and $\alpha/s$ ($s$ is the unit cell area). These properties are calculated with the  expressions derived in this article and using the parameters of first-principles calculation in Table 1 of Ref.\cite{Sohier:2017} (see text).  }
		\begin{tabular}{c c c c c c c c c c}
			\hline
			\hline 
			ML  & $\chi_e$ ($\AA$) & $\chi_0$ ($\AA$) & $\lvert e_B\rvert$ ($e$) & $\lvert e_1\rvert$ ($e$) &   $K_1$ (eV/$\AA^2$) & $c_l$ (km/s)  & $\alpha$ ($\AA^3$) & $\alpha_f$ ($\AA^3$)  &  $\alpha/s$ ($\AA$)  \\  \hline
			 MoS$_2$  & 7.40 & 7.48 & 1.00 &  0.22  &  9.957 & 1.85  & [5.25,10.54] &  18.55 & [0.61,1.22]  \\ 
			 MoSe$_2$ & 8.47 & 8.75 & 1.76 &  0.36  &  8.673 & 4.61  & [6.02,12.18] &  20.52 & [0.64,1.29] \\ 
		    MoTe$_2$ & 11.06 & 12.07 & 3.10 &  0.54  & 6.891 & 13.34 & [7.41,15.20] &  23.71 & [0.69,1.42]  \\ 
			 WS$_2$  & 6.68 & 6.70 & 0.48 &  0.12  &  10.481 & 0.37  & [5.18,10.31] &  16.92 & [0.60,1.20]  \\ 
			 WSe$_2$ & 7.75 & 7.86 & 1.15 &  0.26  &  9.053  & 1.60  & [5.91,11.87] &  18.89  & [0.63,1.27]   \\ 
			 hBN    & 1.22 & 1.67  & 2.71 &  0.46  & 56.495  & 37.10 & [1.43,1.98] &  4.0    
&  [0.26,0.37]  \\ 
			\hline
			\hline 
		\end{tabular}			
	\end{center}			     
\end{table*}

\newpage

\begin{figure}

\caption
{(Color online) Crystal structure of monolayer MoS$_2$. (a) Top view on the hexagonal lattice of the MoS$_2$ monolayer with the Bravais lattice vectors $\mathbf{a}_1$ and $\mathbf{a}_2$, which is similar to the lattice structure of monolayer hexagonal BN.  (b) Side view on the MoS$_2$ monolayer, showing the coordination environment of a Mo atom surrounded by six nearest-neighbour S atoms. The vertical separation between the two S layers is $d$. 
}
\label{fig1}
\vspace*{10mm}

\caption
{(Color online) Longitudinal optical (LO) phonon dispersion relations of (a) monolayer MoS$_2$ and (b) monolayer MoTe$_2$ from the rigid ion model (RIM) and the polarizable ion model (PIM), calculated with or without accounting for local field effects (LFEs) (see text). For the PIM without LFEs, the lower bound $\alpha_l$ and upper bound $\alpha_u$ values of the unit-cell atomic polarizability $\alpha$ in Table~\ref{table:1} are used. Here the wavevector is in units of $\lvert \Gamma-K\rvert$, the distance between the $\Gamma$ and $K$ points in the Brillouin zone. }
\label{fig2}
\vspace*{5mm}

\end{figure}

\newpage

\begin{figure}
\includegraphics*[width=15cm]{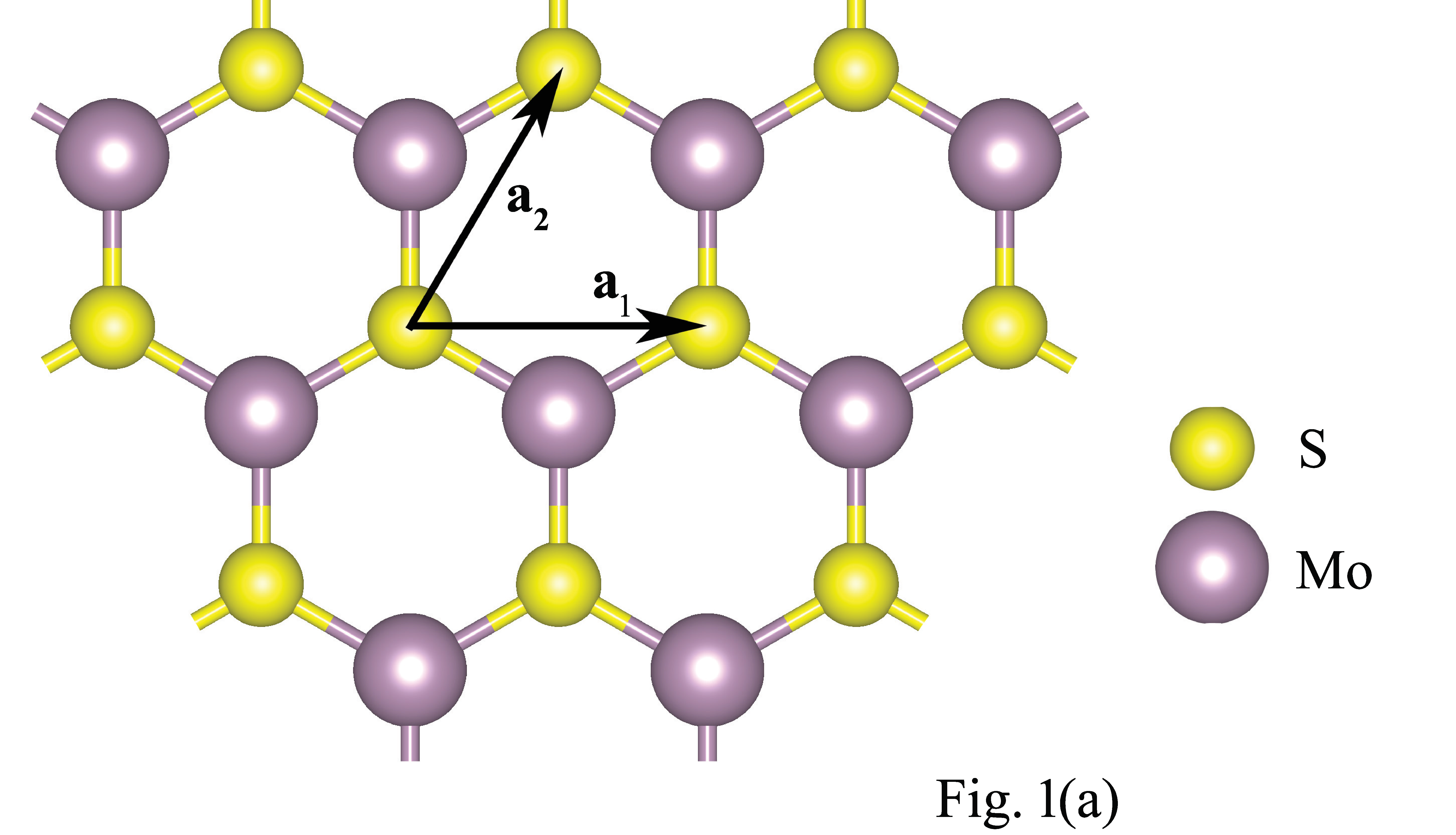}
\end{figure}

\newpage

\begin{figure}
\includegraphics*[width=15cm]{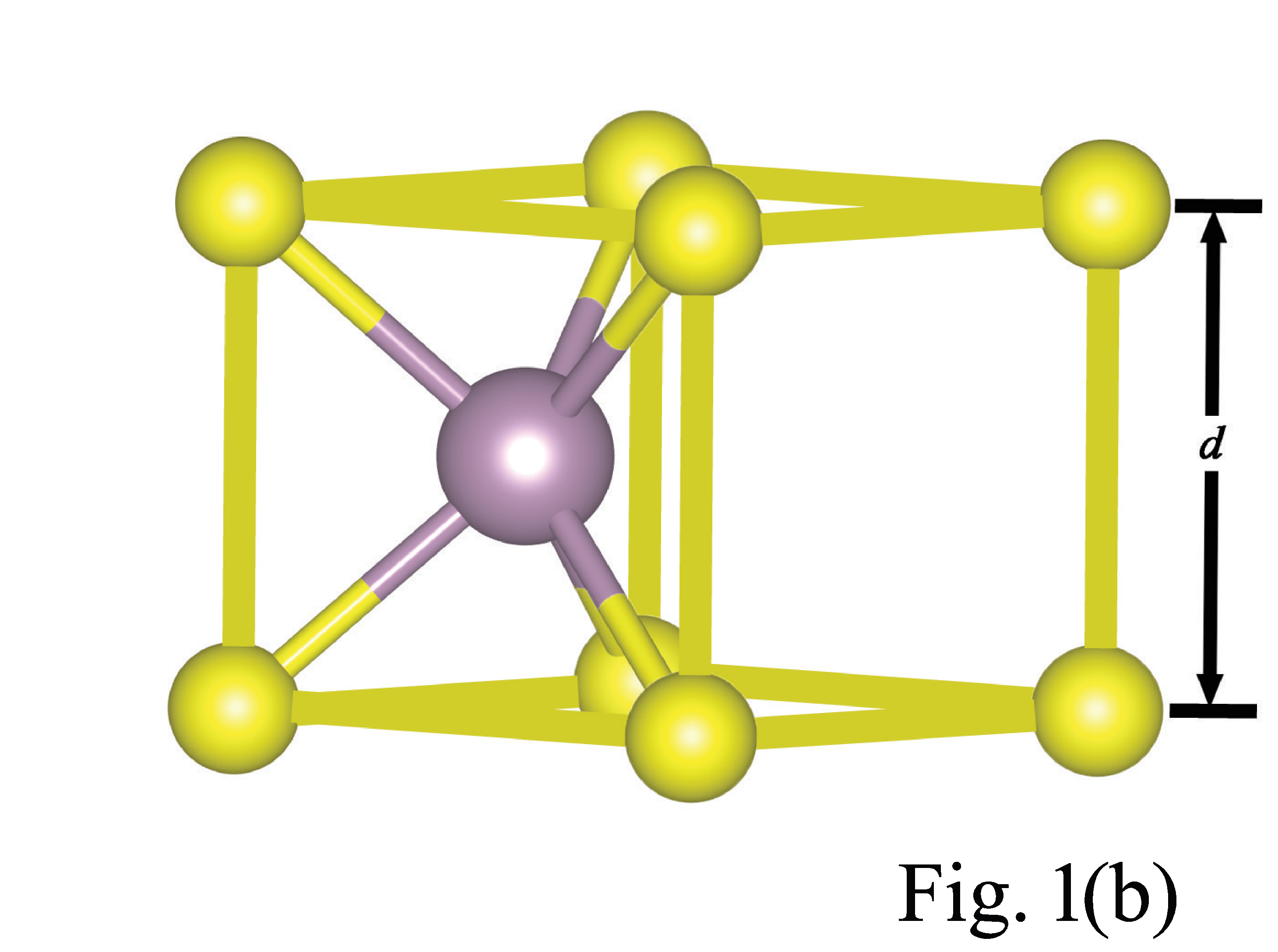}
\end{figure}

\newpage

\begin{figure}
\includegraphics*[width=15cm]{fig2.eps}
\end{figure}

\end{document}